\documentclass{emulateapj}

\newcommand{\cote}{C\^{o}t\'{e}\ }
\newcommand{\jordan}{Jord\'{a}n\ }
\newcommand{\hasegan}{Ha{\c s}egan\ }
\newcommand{\etal}{et~al.\ }
\slugcomment{Accepted for publication in the Astrophysical Journal}

\shorttitle{Diffuse Star Clusters in Early-type Galaxies}
\shortauthors{Peng et al.}

\begin{document}


\title{The ACS Virgo Cluster Survey XI. \\ 
  The Nature of Diffuse
  Star Clusters in Early-Type Galaxies\altaffilmark{1}}


\author{Eric W. Peng \altaffilmark{2}}
%\email{Eric.Peng@nrc-cnrc.gc.ca}

\author{Patrick C\^{o}t\'{e} \altaffilmark{2}}
%\email{Patrick.Cote@nrc-cnrc.gc.ca}

\author{Andr\'{e}s Jord\'{a}n \altaffilmark{3,4}}
%\email{ajordan@eso.org}

\author{John P. Blakeslee \altaffilmark{5,6}}
%\email{jpb@pha.jhu.edu}

\author{Laura Ferrarese \altaffilmark{2}}
%\email{Laura.Ferrarese@nrc-cnrc.gc.ca}

\author{Simona Mei \altaffilmark{5}}
%\email{smei@pha.jhu.edu}

\author{Michael J. West \altaffilmark{7}}
%\email{westm@hawaii.edu}

\author{David Merritt \altaffilmark{8}}
%\email{merritt@astro.rit.edu}

\author{Milos Milosavljevi\'{c} \altaffilmark{9,10}}
%\email{milos@tapir.caltech.edu}

%\and

\author{John L. Tonry \altaffilmark{11}}
%\email{jt@ifa.hawaii.edu}

\altaffiltext{1}{Based on observations with the NASA/ESA {\it Hubble
    Space Telescope} obtained at the Space Telescope Science Institute,
    which is operated by the Association of Universities for Research in
    Astronomy, Inc., under NASA contract NAS 5-26555.}
\altaffiltext{2}{Herzberg Institute of Astrophysics, 
  National Research Council of Canada, 
  5071 West Saanich Road, Victoria, BC  V9E 2E7, Canada; 
  Eric.Peng@nrc-cnrc.gc.ca, Patrick.Cote@nrc-cnrc.gc.ca, 
  Laura.Ferrarese@nrc-cnrc.gc.ca}
\altaffiltext{3}{European Southern Observatory, 
  Karl-Schwarzschild-Str. 2, 85748
  Garching bei M\"{u}nchen, Germany; ajordan@eso.org}
\altaffiltext{4}{Astrophysics, Denys Wilkinson Building, University of
  Oxford, 1 Keble Road, OX1 3RH, UK}
\altaffiltext{5}{Department of Physics and Astronomy, 
  Johns Hopkins University, Baltimore, MD 21218, USA;
  jpb@pha.jhu.edu, smei@pha.jhu.edu}
\altaffiltext{6}{ Department of Physics and Astronomy, 
  Washington State University, Pullman, WA 99164, USA}
\altaffiltext{7}{Department of Physics and Astronomy, University of Hawaii, 
  Hilo, HI 96720, USA; westm@hawaii.edu}
\altaffiltext{8}{Department of Physics, Rochester Institute of Technology,
Rochester, NY 14623-5604, USA; merritt@astro.rit.edu}
\altaffiltext{9}{Theoretical Astrophysics, California Institute of 
  Technology, 
  Mail Stop 130-33, Pasadena, CA 91125, USA; milos@tapir.caltech.edu}
\altaffiltext{10}{Sherman M. Fairchild Fellow}
\altaffiltext{11}{Institute for Astronomy, University of Hawai'i, 2680 Woodlawn
  Drive, Honolulu, HI 96822, USA; jt@ifa.hawaii.edu}
% Mark off your abstract in the ``abstract'' environment. In the manuscript

\begin{abstract}

We use HST/ACS imaging of 100 early-type galaxies in the ACS Virgo
Cluster Survey to investigate the nature of diffuse star clusters (DSCs).
Compared to globular clusters (GCs), these star clusters have
moderately low luminosities ($M_V>-8$) and
a broad distribution of sizes ($3<r_h<30$~pc), but
they are principally characterized by their low mean surface brightnesses
which can be more than three 
magnitudes fainter than a typical GC ($\mu_g > 20\ {\rm mag\
  arcsec}^{-2}$).  The median colors of diffuse 
star cluster systems are red, $1.1<g-z<1.6$, which is redder
than metal-rich GCs and often as red as the galaxy itself.  Most DSC
systems thus have mean ages older than
5~Gyr or else have super-solar metallicities implying that 
diffuse star clusters are likely to be 
long-lived, surviving for significant fraction of a Hubble
time.  We find that 12 galaxies in our sample contain a
significant excess of diffuse star cluster candidates.  Nine of 
them are morphologically classified as lenticulars (S0s),
and five of them visibly contain dust.  
We also find a substantial population of DSCs in
the halo of the giant elliptical M49, associated with the companion
galaxy VCC~1199.  
Most DSC systems appear to be both aligned with the galaxy light and 
associated with galactic disks, but
at the same time many lenticular galaxies do not host substantial DSC
populations, and environment and clustercentric radius do not appear to be
good predictors of their existence.  
Diffuse star clusters in our sample share similar
characteristics to those identified in other nearby lenticular, spiral,
and dwarf galaxies.  
Unlike luminous GCs, whose sizes are constant
with luminiosity, DSCs are bounded at the bright end by an
envelope of nearly constant surface brightness.  
We suggest that populations of diffuse star clusters preferentially
form, survive, and coevolve with galactic disks.  Their properties are
broadly consistent with those of merged star cluster complexes, and
we note that despite being 3--5
magnitudes brighter than DSCs, ultra-compact dwarfs have similar surface
brightnesses.  The closest Galactic analogs to the DSCs are the old open
clusters.  We suggest that if a diffuse star cluster population did
exist in the disk of the Milky Way, it would be very difficult to find.

\end{abstract}


%% subject header.  Objects should be in the appropriate "individual"
%% headers (e.g. quasars: individual, stars: individual, etc.) with the
%% additional provision that the total number of headers, including each
%% individual object, not exceed six.  The \objectname{} macro, and its
%% alias \object{}, is used to mark each object.  The macro takes the object
%% name as its primary argument.  This name will appear in the paper
%% and serve as the link's anchor in the electronic edition if the name
%% is recognized by the data centers.  The macro also takes an optional
%% argument in parentheses in cases where the data center identification
%% differs from what is to be printed in the paper.



\bigskip
\section{Introduction}

In the Milky Way, star clusters have long been divided into two
distinct classes---ancient globular clusters (GCs) that populate the halo
and bulge, and the young and less massive open clusters that reside in the
disk.  Globular clusters appear to be a ubiquitous stellar population,
present in galaxies of nearly all morphological types, luminosities, and
star formation histories.  Studies of extragalactic GCs over the past decade,
especially with the Hubble Space Telescope ($HST$), have revealed
properties that are uniform or slowly varying across galaxies---their
distributions of size, color, and luminosity.

Some observations of nearby spiral, irregular, and lenticular
galaxies, however,
have revealed a wider array of star cluster characteristics, all of
which are ultimately windows onto the galactic environment within which
they form and reside.  In addition to GCs, these galaxies also contain
star clusters for which there are  no Galactic analogs, for example the
young ``populous'' 
star clusters in the Magellanic Clouds and M33 that are intermediate in
mass between open and globular clusters and do not fit the Galactic
dichotomy (e.g.\ Chandar, Bianchi, \& Ford 1999).
In M101 and NGC~6946, $HST$ imaging reveals a larger than expected
number of faint, compact, old star 
clusters that are similar to GCs but have numbers that continue
to rise at luminosities fainter than the typical turnover value of the
globular cluster luminosity function (Chandar, Whitmore, \& Lee 2004).  

$HST$ imaging has also uncovered a different population of faint and
unusually extended old star clusters in two lenticular galaxies NGC~1023
and NGC~3384 (Larsen \& Brodie 2000; Larsen \etal 2001).  
These clusters (which the authors
call ``faint fuzzies'') appear to be moderately metal-rich 
(${\rm [Fe/H]}\sim-0.6$), have old ages ($>7$~Gyr) and are both
spatially and kinematically associated with the stellar disks of the
lenticular galaxies in which they were serendipitously discovered
(Brodie \& Larsen 2002).  However, they find that neither the S0 galaxy
NGC~3115 nor the elliptical NGC~3379 seems to possess a similar population.

Extended clusters that appear atypical for the Milky Way have also been
identified in M31 (Huxor \etal 2005) and M33 (Chandar \etal 1999), and
appear particularly numerous in M51 (Chandar \etal 2004).  Sharina,
Puzia, \& Makarov 
(2005) also find a large fraction star clusters that are fainter and
more extended than typical GCs in nearby dwarf galaxies
($D=2$--$6\ {\rm Mpc}$), both dSphs and dIs.  
van den Bergh (2005a) combine
the Sharina \etal (2005) data with that for low-luminosity galaxies in
the Local Group find that dwarf galaxies have an excess of faint old star
clusters as compared to giant galaxies.  Most recently, G\'{o}mez \etal
(2005) have found that GCs in the peculiar elliptical galaxy NGC~5128
(Cen~A) have sizes and ellipticities that extend to higher values than
their Milky Way counterparts.

What are the conditions that give rise to these star
clusters?  The constituents of cluster systems are a direct reflection of
the variegated formation histories of the parent galaxies---the 
numbers and masses of star clusters is proportional to the star
formation rate density (Larsen \& Richtler 2000), 
while the survival of these clusters is dependent
on their compactness, their orbits, their age, and the local
gravitational potential (e.g.\ Fall \& Rees 1977; Zhang \& Fall 2001).  By
investigating the nature of star cluster 
systems for which there are no or few known Local Group analogs, we are
studying a much wider array of star forming environments and galactic
evolutionary histories.

At present, it is unclear whether the star clusters in these galaxies
are all the result of the same phenomena, or are formed and destroyed
through many different processes.  Some appear metal-poor (M31; Huxor
\etal 2005) and some
metal-rich (NGC~1023, 3384). Many appear associated with galactic disks, but at
least a few in M33 are not.  Fellhauer \& Kroupa (2002) have proposed
that these star 
clusters can be formed through the merging
of star cluster complexes, and may be related to the hierarchical
nature of star formation in disks (Bastian \etal 2005).
Some of these clusters also may not be all
that different from Milky Way GCs or open clusters.  
Observationally, however, they share
one common property---the accumulating evidence for this newfound
diversity in extragalactic 
star cluster systems is a direct result of our ability to study fainter
and lower surface brightness star clusters in other galaxies.  Being
able to resolve the sizes of these clusters is essential
to all these studies.  

In this investigation, we present a study of low surface brightness or
``diffuse'' star cluster (DSC) populations in the ACS Virgo
Cluster Survey (ACSVCS, \cote \etal 2004, Paper~I).  
The ACSVCS is a program to image 100 early-type galaxies in the Virgo
cluster with the HST Advanced Camera for Surveys (ACS).  This survey
is designed to study the star cluster systems of these galaxies in a
deep and homogeneous manner.  Moreover, the high spatial resolution afforded by
HST/ACS enables us to measure sizes for all star clusters (\jordan
\etal 2005).  As such, the survey is well-designed to
determine the properties of star clusters that have lower surface
brightnesses than typical globular clusters, and to determine their
frequency in early-type galaxies.

\section{Observations and Data}
\subsection{The Sample}

A full description of the ACS Virgo Cluster Survey is provided in
Paper~I.  Here, we briefly describe the sample and observations.  The
ACSVCS is a program which obtained HST/ACS images in the F475W ($g$)
and F850LP ($z$) filters of 100 early-type galaxies from the Virgo
Cluster Catalog (VCC) of Binggeli, Sandage, \& Tammann (1985,
hereafter BST85).  These galaxies were selected to be certain or
probable members of the Virgo cluster excluding the Southern
Extension, have $B_T<16$, and morphologically classified as E, S0, dE,
dE,N, dS0, or dS0,N.  From these 163 galaxies, we were required to
trim to a sample of 100 because of time limitations, eliminating 63
galaxies that were S0 or dwarfs with uncertain or disturbed morphologies,
significant dust, lacking a visible bulge, or had prior HST/WFPC2 imaging.
We note that despite this culling, this early-type sample is magnitude
limited for the brightest 26 galaxies with $B_T<12.15$ ($M_B <
-18.94$), and that 38 of the 100 galaxies in our sample are classified
as S0, E/S0, or S0/E (see Paper~I for the full list of galaxies and
their properties).  For the purposes of this paper, we adopt a
distance to the Virgo Cluster of $D=16.5~{\rm Mpc}$ with a distance
modulus of $31.09\pm0.03$~mag from Tonry \etal (2001), corrected by
the final results of the Key Project distances (Freedman \etal 2001;
see also discussion in Mei \etal 2005).

All observations were taken with the Wide Field Channel (WFC) of the ACS.
The WFC has two $2048\times4096$ CCDs butted together with a scale of
$0\farcs049\ {\rm pixel}^{-1}$ and a field of view of
$202\arcsec\times202\arcsec$.  At the distance of the Virgo Cluster,
this translates into 4~pc~pixel$^{-1}$ and a 
$16.2\ {\rm kpc} \times16.2\ {\rm kpc}$ field of view.  Exposure
times in the F475W and F850LP filters totaled 720~sec and 1210~sec,
respectively.

\subsection{Data Reduction}

Our ACS/WFC images were reduced using a dedicated pipeline which is
described in \jordan \etal (2004a,b, Papers II and III).  The science images
are produced by combining them and cleaning them of cosmic rays using
the Pyraf task {\it multidrizzle} (Koekemoer \etal 2002).  In order to
detect star clusters, it is first important to remove the light from
the galaxy.  We subtract a model of the galaxy in each filter and
subsequently iterate with the source detection program
SExtractor (Bertin \& Arnouts 1996) in order to mask objects and remove any
residual background.  We do our final object detection using estimates
of both the image noise and the noise due to surface brightness
fluctuations.  Objects are accepted into our catalog only if
they are detected in both filters.  Very bright or elongated objects
are rejected to eliminate obvious foreground stars and background
galaxies.  For the remaining objects we use the program KINGPHOT
(\jordan \etal 2005) to measure magnitudes and King model parameters
for all candidate star clusters.  KINGPHOT fits King model surface
brightness profiles convolved with the point spread function (PSF)
appropriate for each object in each filter.  For total magnitudes, we
integrate the flux of the best-fit PSF-convolved model to the limit of
the PSF and apply an aperture correction that is dependent on the
fitted half-light radius, $r_h$ (see description in Peng \etal 2005,
Paper~IX).
Magnitudes and colors are corrected for foreground extinction using
the reddening maps of Schlegel, Finkbeiner, \& Davis (1998) and
extinction ratios for the spectral energy distribution of a G2 star
(Paper II; Sirianni \etal 2005).  For the purposes of this paper, we use
$g$ and $z$ to mean the ACS magnitudes $g_{475}$ and $z_{850}$.
 
%%%%%%%%%%%%%%%%%%%%%%%%%%%%%%%%%%%%%%%%%%%%%%%%%%%%%%%%%%%%%%%%%%%%%%%%
\begin{figure}
%\epsscale{.80}
\epsscale{1.2}
\plotone{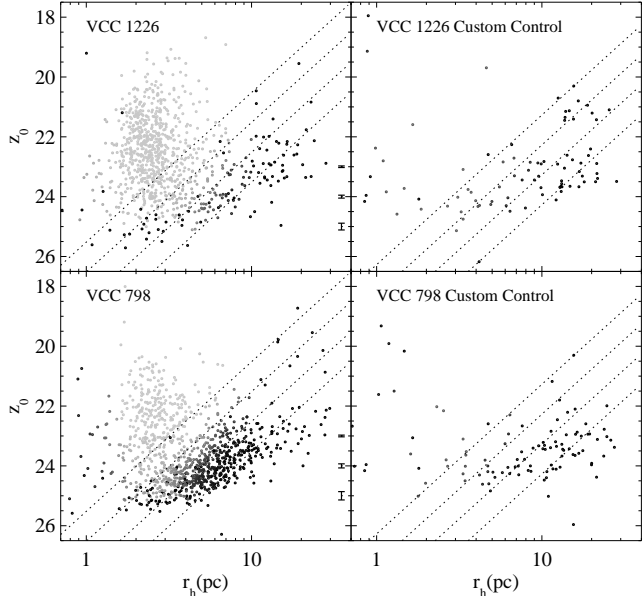}
\caption{Size-magnitude selection 
  diagrams and control fields for VCC~1226 (M49, top) and 
  VCC~798 (M85, bottom).  Gray points with a mean half-light radius of
  $r_h=2.6$~pc are objects that likely belong to the traditional
  population of globular clusters.  Black points represent objects
  classifies as likely to be contaminants.  The left panels show the
  data for the the galaxy fields, and the right panels show what is
  expected from the custom control field for each galaxy.  Diagonal
  dotted lines are lines of constant mean surface brightness, 
  $\mu_z = 18,\, 19,\, 20,\, 21\ {\rm mag\ arcsec}^{-2}$ (left to right).
  Notice that
  VCC~798, despite having a smaller GC population than VCC~1226, 
  has many more objects
  in the ``contaminant'' locus than would be expected from the control
  fields.  These lower surface brightness objects are the star clusters
  of interest.  Error bars represent the median
  photometric error at those magnitudes. Median errors in $r_h$ for 
  those bins are 0.26, 0.57, and 1.22~pc, respectively.  \label{fig:m49m85}}
\end{figure}
%%%%%%%%%%%%%%%%%%%%%%%%%%%%%%%%%%%%%%%%%%%%%%%%%%%%%%%%%%%%%%%%%%%%%%%%

\subsection{Control Fields}

A critical issue in any extragalactic star cluster study that
pushes to faint apparent magnitudes is contamination from background
galaxies.  In Peng \etal (2005), we described an approach where we
run our pipeline on 17 blank, high-latitude control fields taken
from the ACS Pure Parallel Program (GO-9488 and 9575).  Even with
equivalent exposure time, however, it is important to remember that
blank fields will always detect fainter objects than our ACSVCS images
because the bright galaxy in our science images necessarily creates a
spatially varying detection efficiency that is a function of the
background light.  We address this issue by creating ``custom''
control samples for each galaxy.  To do this, we run the detection
on each control field as if that particular Virgo galaxies were in the
foreground.  In this way, we are able to create a true comparison
sample tailored to each ACSVCS galaxy.

\subsection{Star Cluster Selection}

%%%%%%%%%%%%%%%%%%%%%%%%%%%%%%%%%%%%%%%%%%%%%%%%%%%%%%%%%%%%%%%%%%%%%%%%%%
\begin{figure}
\epsscale{1.2}
\plotone{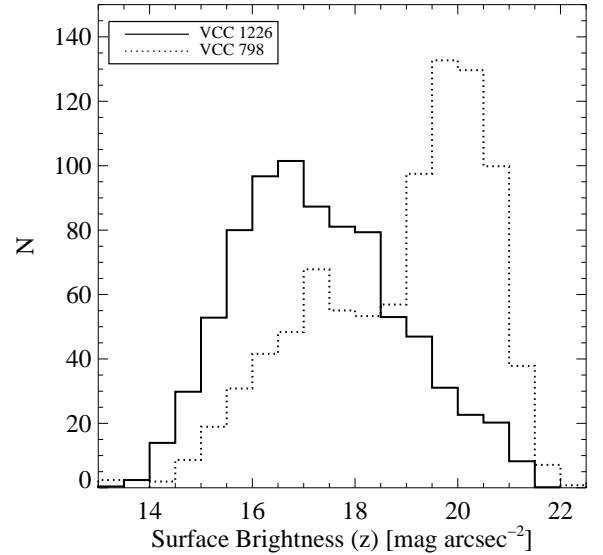}
\caption{Histograms of $z$ surface brightness for objects in VCC~1226
  (solid) and VCC~798 (dotted).  Expected background galaxy contamination 
  as measured in the custom control fields has been subtracted.
  As in Figure~\ref{fig:m49m85}, compared to VCC~1226, VCC~798 has a
  large population of star clusters with lower surface brightnesses
  ($z\gtrsim19\ {\rm mag\ arcsec}^{-2}$).  
  \label{fig:sbhist}}
\end{figure}
%%%%%%%%%%%%%%%%%%%%%%%%%%%%%%%%%%%%%%%%%%%%%%%%%%%%%%%%%%%%%%%%%%%%%%%%%%

In Paper~IX, we used our carefully constructed control samples to
optimally identify globular clusters.  Our ability to measure sizes
for objects provides leverage in the size-magnitude plane to separate
globular clusters from other objects.  The control fields allow us to
localize the contaminating population in this plane (mostly background
galaxies).  Typical GCs have median half-light radii of $r_h\sim3$~pc
(e.g.\ \jordan \etal 2005) and follow a Gaussian-like luminosity
function.  On the other hand, the contaminant population consists of
point sources at all magnitudes (foreground stars), and a background
galaxy population that consists of faint, compact galaxies as well as
brighter, more extended galaxies.  This is demonstrated in the top
panel of Figure~\ref{fig:m49m85} which shows the $r_h$--$z_0$ diagram
for the brightest galaxy in our sample, M49 (VCC~1226), as well as
its custom control field.  We assign each object in this diagram a
``GC probability'' based on its position relative to the loci of GCs
(assumed) and contaminants (measured).  Objects with GC probabilities
greater than 0.5 are included in our GC sample.  We describe the
details of the algorithm in Paper~IX and \jordan \etal (2006).

It is important to note, however, that the GC and contaminant loci
have the highest degree of overlap for objects that have low surface
brightnesses ($\mu_z\gtrsim 19\ {\rm mag\ arcsec}^{-2}$),
where $\mu_z$ is the mean surface brightness within the half-light radius.
Therefore, faint or extended star clusters are not easily separated from
background galaxies on the basis of magnitude and size alone.  While
in Paper~IX, we purposely focused our study on the ``traditional'' GC
population, the focus of this paper is to investigate the existence
and properties of star clusters that have lower surface brightnesses
than typical GCs, and that are more easily confused with background objects.
Therefore, for the purposes of this paper, we define this sample as
all objects with GC probabilities less than 0.2 and half-light radii
$r_h>4$~pc.  We set this strict
criteria so that we may avoid contamination from the sample of
traditional GCs.  While the exact numbers for this cutoff are not
critical, we chose these values because they are inclusive of many
objects yet the background corrected numbers for this cut is still close
to zero for most galaxies (i.e. they are uncontaminated by GCs even for
the massive ellitpicals).

%%%%%%%%%%%%%%%%%%%%%%%%%%%%%%%%%%%%%%%%%%%%%%%%%%%%%%%%%%%%%%%%%%%%%%%%%%
\begin{figure}
\epsscale{1.2}
\plotone{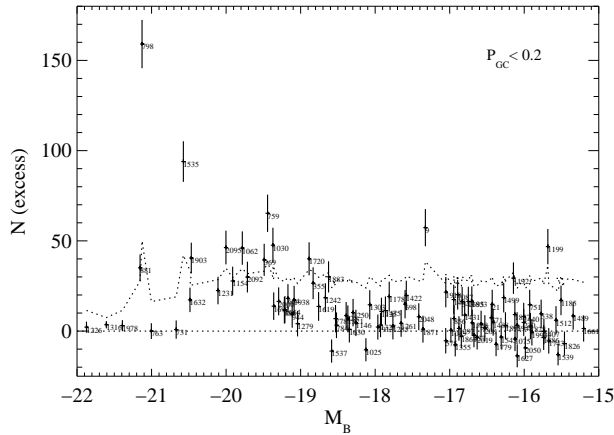}
\caption{Number of ``non-GCs'' versus galaxy absolute B magnitude.
  We show the number of objects with GC probabilities less than 0.2
  and $r_h > 4$~pc that are in excess of what is expected for the
  background as estimated from the control fields.  Numbers alongside
  the points are the VCC numbers for each galaxy.  The dotted line
  represents a three-sigma value in the excess distribution.  Note
  that while VCC~798 is the largest outlier, a number of galaxies,
  notably around $M_B=-20$, have a significant excess of objects in
  this region of parameter space.
  \label{fig:nvsmb}}
\end{figure}
%%%%%%%%%%%%%%%%%%%%%%%%%%%%%%%%%%%%%%%%%%%%%%%%%%%%%%%%%%%%%%%%%%%%%%%%%%

\section{Results}
\subsection{The Existence and Frequency of Diffuse Star Cluster Populations}

We can test each galaxy for the existence of a population of diffuse
star clusters (DSCs) by comparing the total number of objects in
this region of the size-magnitude plane with that expected from the
control fields.  When we do this, we find that some galaxies have a
clear excess of objects in this locus.  Figure~\ref{fig:m49m85} 
shows a comparison
between the normal giant elliptical VCC~1226 (M49) and the S0 galaxy
VCC~798 (M85).  Diagonal dotted lines are loci of constant $z$
surface brightness, $\mu_z$.
While the number of objects with low GC probabilities
($<0.2$) in VCC~1226 is consistent with that in the custom control sample,
the number of these objects in VCC~798 --- those with 
$\mu_z\gtrsim19\ {\rm mag\ arcsec}^{-2}$ --- is well in excess of the background.  

This is further illustrated in Figure~\ref{fig:sbhist} where we show
the background subtracted 
$\mu_z$ distributions for all star cluster candidates in
VCC~1226 and 798.  VCC~798 has a large population of star clusters at
lower surface brightnesses, and   
of the 100 galaxies in the ACSVCS sample, VCC~798 has the largest number
of DSC candidates.
Note that the turnover of the surface
brightness distribution at $\mu_z\sim20$ is due to incompleteness
and not an intrinsic property of the objects themselves.  

In Figure~\ref{fig:nvsmb}, 
we show the number of ``non-GCs'' that are in excess of
the expected background for the entire ACSVCS sample.  Most of the
galaxies have almost no excess population of
objects.  This shows that our custom control fields are in fact
providing reasonable estimates of the background.  It is also the case
that galaxies with both large and small numbers of globular clusters
can have excesses near zero, showing that our chosen criterion of
$P_{GC}<0.2$ is stringent enough to eliminate most of the traditional
GC sample from appearing our sample of potential DSCs.

\begin{figure}
\epsscale{1.2}
\plotone{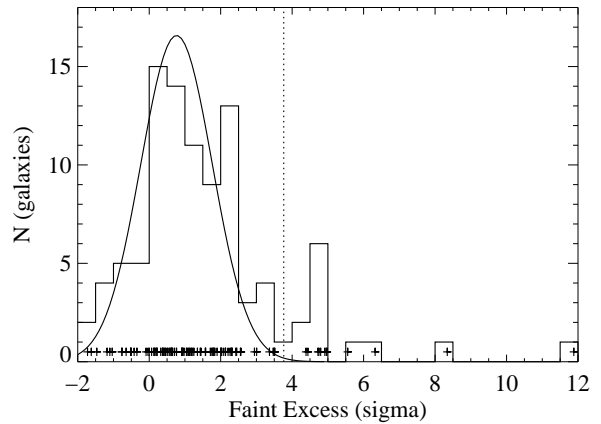}
\caption{Distribution of ``non-GC'' excess in units of sigma over the
  background.  Data is the same as for Figure~\ref{fig:nvsmb}.
  The expected width of the distribution as determined by
  counting statistics and cosmic variance is represented by the
  Gaussian curve.  While the mean has been adjusted to fit the galaxy
  data, its width has not.  Pluses along the bottom represent the
  position of individual galaxies.  Notice that the distribution is
  largely Gaussian, but has a tail of objects to the right.  The
  dotted line represents $3\sigma$ above the mean, and separates
  ``normal'' galaxies from those that have an excess of faint,
  extended objects.
  \label{fig:fxhist}}
\end{figure}

In the figure, we also plot a dotted line that represents the point at
which the number of objects are in excess of the background at a level
greater than $3\sigma$ above the mean.  
The expected variance for each galaxy is
determined using the 17 control fields.  We count the number of
objects that meet the selection criteria in each field and determine
the variance of this number over all the control fields.  This is then
added to the variance expected from Poisson counting statistics in the
science field itself.  Figure~\ref{fig:fxhist} shows a histogram of
galaxies, binned by the 
diffuse source excess in units of $\sigma$.  While the
mean of the distribution is slightly shifted (around 0.8$\sigma$), the
width of the distribution is fixed by our error bars, and thus the
fact that a unit Gaussian is a reasonable match to the distribution
shows that our error bars are also reasonable.

%%%%%%%%%%%%%%%%%%%%%%%%%%%%%%%%%%%%%%%%%%%%%%%%%%%%%%%%%%%%%%%%%%%%%%%%%%
%%%%%%%%%%%%%%%%%%%%%%%%%%%%%%%%%%%%%%%%%%%%%%%%%%%%%%%%%%%%%%%%%%%%%%%%%%

\begin{deluxetable*}{rrcccccc}[b]
\tablewidth{0pt}
\tablecaption{Galaxies with $>3\sigma$ Excess of $P_{GC}<0.2$ Sources\label{table:excesstable}}
\tablehead{
\colhead{VCC} & 
\colhead{$M_B$} & 
\colhead{$(g$--$z)_{galx}$} & 
\colhead{Type(VCC)} & 
\colhead{$(g$--$z)_{DSC}$} & 
\colhead{$N_{DSC}$} & 
\colhead{$S_{N,DSC}$} & 
\colhead{$T_{L,DSC}(z)$} \\ 
\colhead{(1)} &
\colhead{(2)} &
\colhead{(3)} &
\colhead{(4)} &
\colhead{(5)} &
\colhead{(6)} &
\colhead{(7)} &
\colhead{(8)} 
}
\startdata
 881 & $-21.2$ &   1.51 & S0/E &   1.27 &   32 &  0.09 &    1.2 \\
 798 & $-21.1$ &   1.32 & S0   &   1.33 &  160 &  0.36 &    4.2 \\
1535 & $-20.6$ & \nodata & S0   &   1.54 &   93 &  0.40 &    4.7 \\
1903 & $-20.5$ &   1.46 & E    &   1.40 &   38 &  0.17 &    2.1 \\
2095 & $-20.0$ &   1.39 & S0   &   1.29 &   48 &  0.34 &    3.6 \\
1062 & $-19.8$ &   1.48 & S0   &   1.48 &   45 &  0.30 &    3.7 \\
 369 & $-19.5$ &   1.47 & S0   &   1.48 &   37 &  0.38 &    5.0 \\
 759 & $-19.4$ &   1.46 & S0   &   1.49 &   65 &  0.55 &    7.0 \\
1030 & $-19.4$ & \nodata & S0   &   1.37 &   47 &  0.52 &    6.1 \\
1720 & $-18.9$ &   1.40 & S0   &   1.43 &   38 &  0.51 &    7.2 \\
   9 & $-17.3$ &   1.06 & dE   &   1.11 &   59 &  4.20 &   46.4 \\
1199 & $-15.7$ &   1.52 & E    &   1.42 &   45 & 24.44 &  352.6 \\
\enddata
\tablenotetext{1}{Number in Virgo Cluster Catalog}
\tablenotetext{2}{Absolute B Magnitude, extinction-corrected, $D=16.5$~Mpc}
\tablenotetext{3}{Galaxy color from Paper VI}
\tablenotetext{4}{Morphological type from the VCC}
\tablenotetext{5}{Median color of DSCs}
\tablenotetext{6}{Number of DSCs}
\tablenotetext{7}{Specific frequency of DSCs: $S_{N,DSC} = N_{DSC}\times 10^{0.4(M_V+15)}$}
\tablenotetext{8}{Specific $z$ luminosity of DSCs: $T_{L,DSC}(z) = L_{z,DSC}/L_{z,galaxy} \times 10^4$}
\end{deluxetable*}

Figure~\ref{fig:fxhist} also shows that there is a significant tail of
galaxies that have very high numbers of low surface brightness sources.
In addition to VCC~798 and VCC~1535 which Figure~\ref{fig:nvsmb} shows 
to have the largest number of sources in excess of the background, 
there are a total of twelve
galaxies that have an excess greater than $3\sigma$.  These galaxies
and some of their properties are listed in
Table~\ref{table:excesstable}.  For the purpose of investigating the
nature of the diffuse star cluster 
populations, we focus this paper on these twelve
galaxies that have $>3\sigma$ excesses of faint, extended sources.
However, we note that a number of galaxies that have less significant
excesses also possibly harbor populations of DSCs.

We show the size-magnitude ($r_h$--$z_0$) diagram for each one of
these twelve galaxies and their custom control fields 
in Figure~\ref{fig:selectionfig}.  The objects that are likely to be
GCs are gray, while objects that would normally be classified as
contaminants are black.  In each case, 
there is an excess of low surface brightness objects over what
one sees in the custom control field.  The galaxies are ordered by
their total $B$ luminosity, and we point out that fainter galaxies
have fainter objects included in their custom control fields, as one
would expect with the lower average background.  It is also clear that
for most galaxies at the distance of the Virgo Cluster, there is a
substantial background population and one must be very careful
drawing conclusions about the existence of DSCs without a careful
treatment of the background.  This is a problem in {\it apparent}
magnitude, so will be less of an issue for more nearby star cluster
systems, and more of problem for more distant star cluster systems.

%%%%%%%%%%%%%%%%%%%%%%%%%%%%%%%%%%%%%%%%%%%%%%%%%%%%%%%%%%%%%%%%%%%%%%%%%%
\begin{figure}
\epsscale{1.2}
\plotone{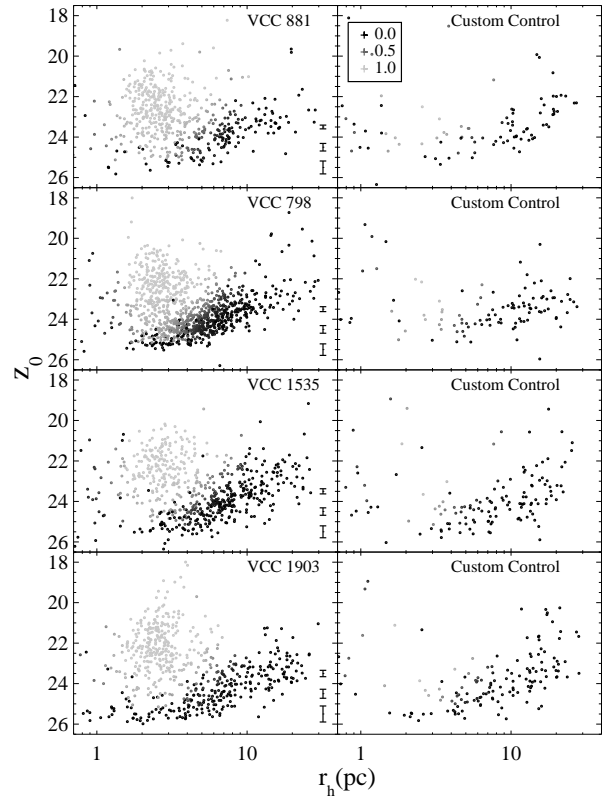}
\caption{Size-magnitude diagrams for the twelve galaxies that have
  significant diffuse source excesses, listed in
  Table~\ref{table:excesstable}.  Using the algorithm described in the
  text, we assign a GC probability based on its position in the $r_h$--$z_0$
  diagram.  Objects are color coded by their probability of being a
  globular cluster.  On the left are the objects in our program
  fields, and on the right are the objects in the custom control fields
  for that galaxy, scaled to a single field.  Each one of these
  galaxies shows at least a 3$\sigma$ excess of diffuse
  sources as compared to the control fields. \label{fig:selectionfig}}
\end{figure}
%%%%%%%%%%%%%%%%%%%%%%%%%%%%%%%%%%%%%%%%%%%%%%%%%%%%%%%%%%%%%%%%%%%%%%%%%%

%%%%%%%%%%%%%%%%%%%%%%%%%%%%%%%%%%%%%%%%%%%%%%%%%%%%%%%%%%%%%%%%%%%%%%%%%%
\begin{figure}
\epsscale{1.2}
\plotone{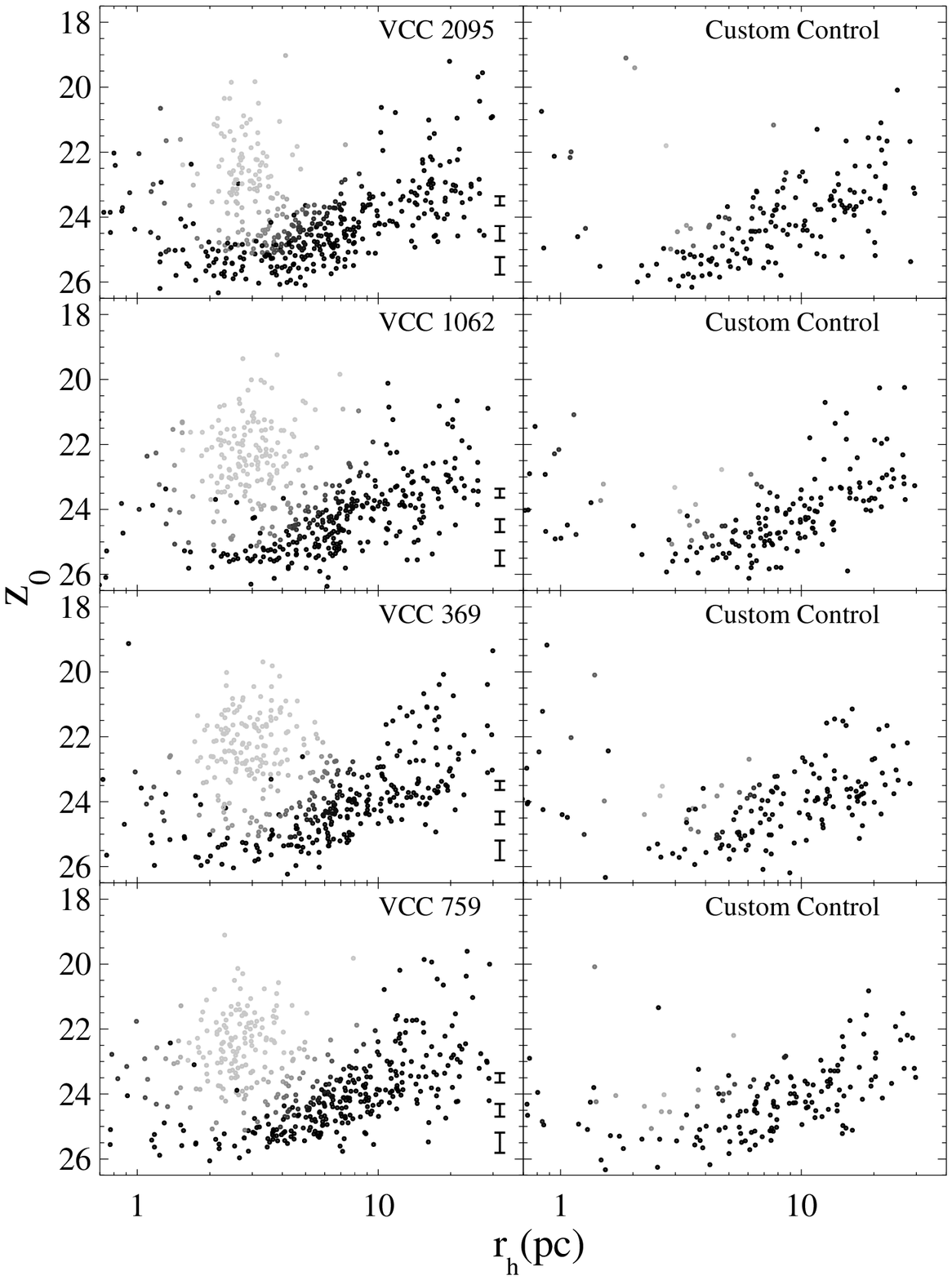}
\figurenum{5}
\caption{continued.  Size-magnitude diagrams for the twelve galaxies that have
  significant diffuse source excesses.}
\end{figure}
%%%%%%%%%%%%%%%%%%%%%%%%%%%%%%%%%%%%%%%%%%%%%%%%%%%%%%%%%%%%%%%%%%%%%%%%%%

%%%%%%%%%%%%%%%%%%%%%%%%%%%%%%%%%%%%%%%%%%%%%%%%%%%%%%%%%%%%%%%%%%%%%%%%%%
\begin{figure}
\epsscale{1.2}
\plotone{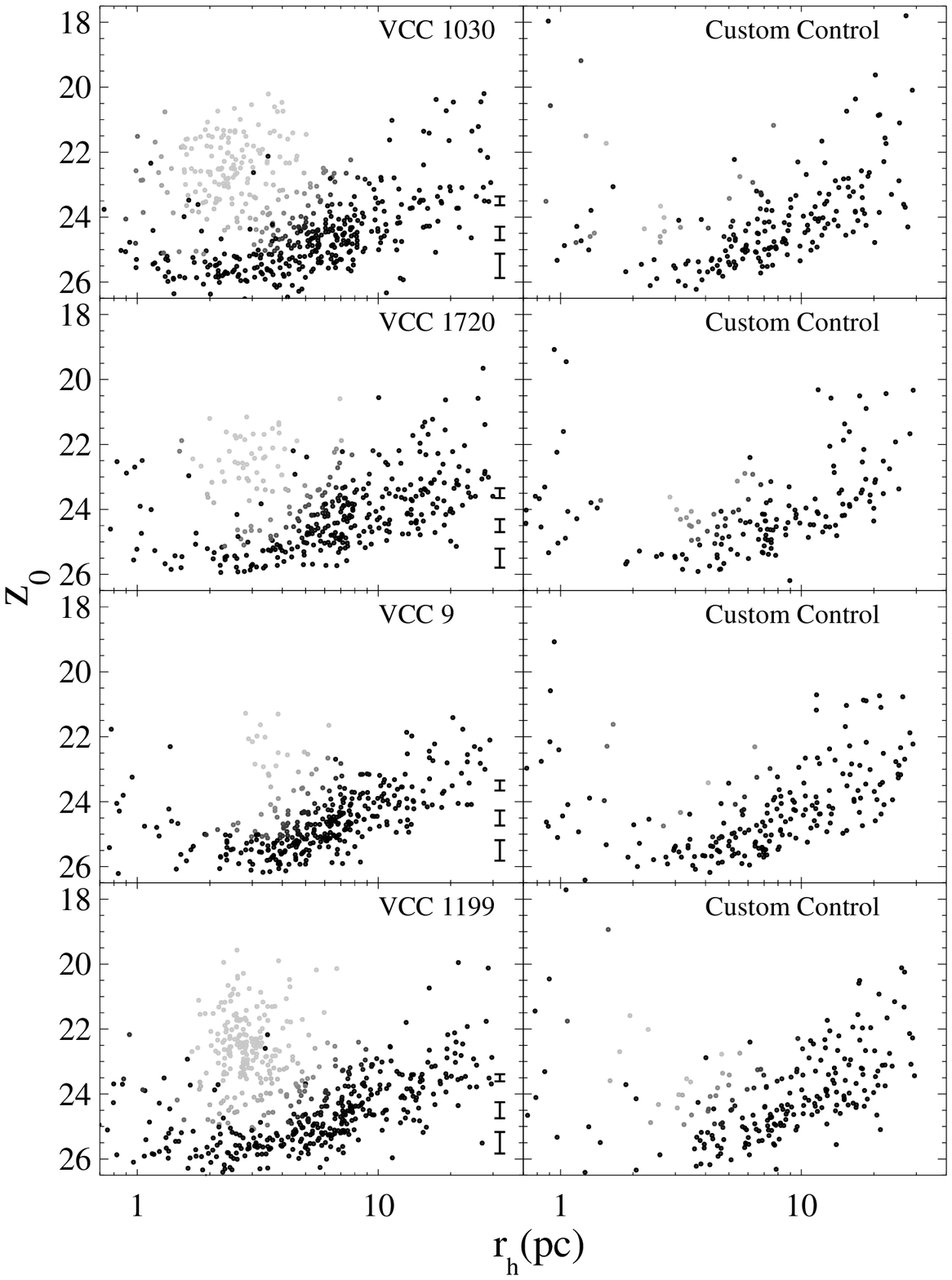}
\figurenum{5}
\caption{continued.  Size-magnitude diagrams for the twelve galaxies that have
  significant diffuse source excesses.}
\epsscale{1.0}
\end{figure}
%%%%%%%%%%%%%%%%%%%%%%%%%%%%%%%%%%%%%%%%%%%%%%%%%%%%%%%%%%%%%%%%%%%%%%%%%%

Of the twelve galaxies that have a significant population of
DSCs, nine of them are classified as S0 or S0/E: VCC~881, 798, 1535,
2095, 1062, 369, 759, 1030, and 1720.  Of the three galaxies that were
not classified as S0 by BST85: VCC~1903 is classified as an E5 in the
RC3 (de Vaucouleurs \etal 1991); VCC~9 is a low surface
brightness galaxy that appears atypical for its luminosity---it is a
dwarf with a ``core'' and is possibly a dI/dE transition object; and
VCC~1199 is a companion of VCC~1226 (M49), only $4\farcm5$ away, and
it is likely that most of the star clusters we see belong to the cluster
system of the nearby giant elliptical.  Five of the galaxies (all of
them S0s) also visibly contain dust---VCC~881, 798, 1535, 759, and 1030.

\subsection{Properties of Diffuse Star Clusters: Colors, Magnitudes, Sizes}
\begin{figure}
\plotone{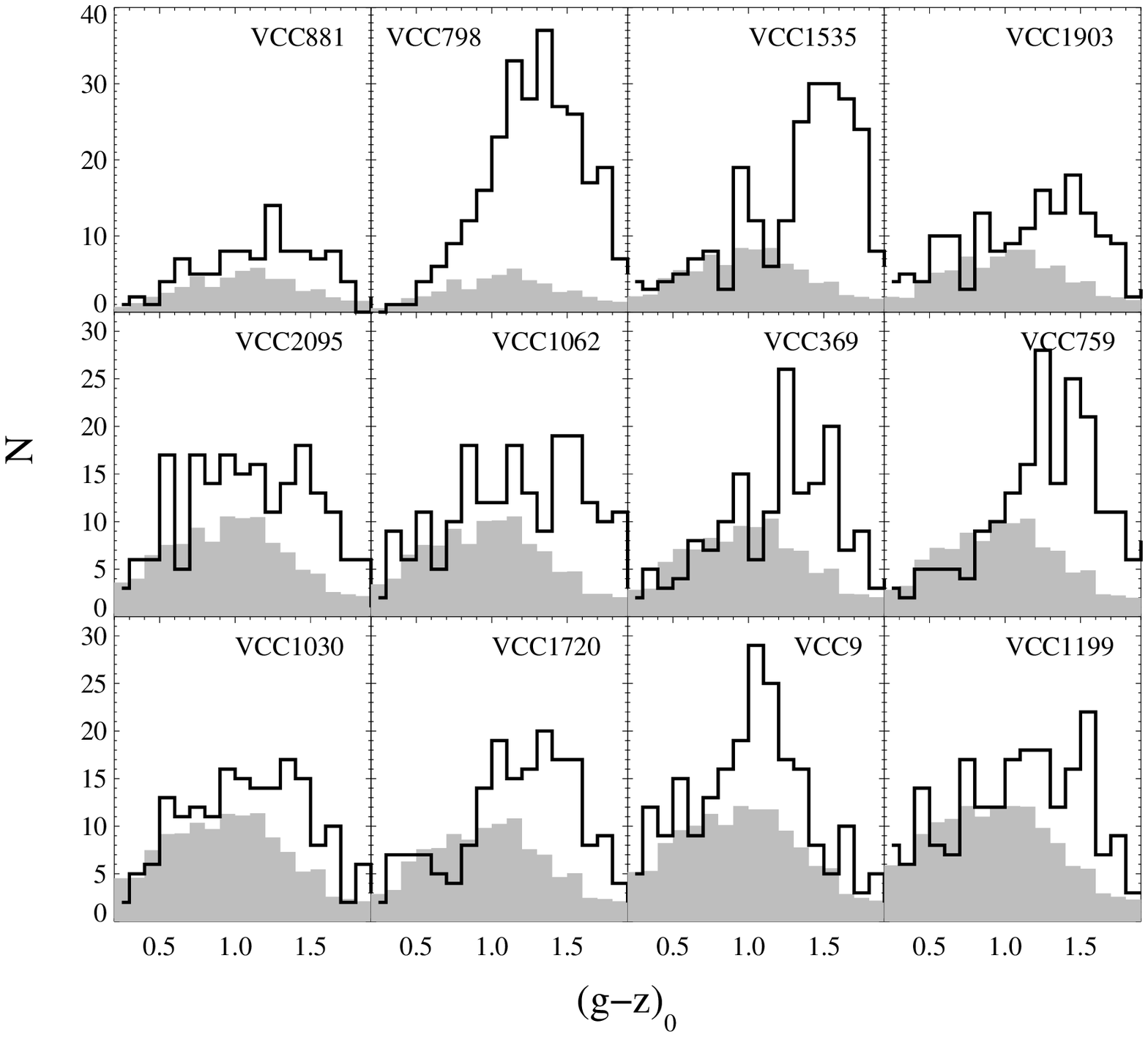}
\caption{Color distributions of diffuse sources (black) compared to
  that for expected background contamination (filled gray) 
  for the 12 galaxies in
  Table~\ref{table:excesstable}.  Galaxies are ordered by decreasing
  $B$ luminosity.  All histograms use 0.1~mag bins.  Notice that the
  colors of the diffuse sources are redder than that the
  typical background contaminant.  While the distribution of
  diffuse sources bluewards of $(g$--$z)_0\sim1.0$ is generally 
  well-matched by
  the control fields, these galaxies clearly harbor an excess of red
  sources. \label{fig:fes_bg}}
\end{figure}

Although the diffuse star clusters 
have sizes and magnitudes similar to the background
galaxy population, their color distributions may be very different.
For all twelve ``excess'' galaxies, we show in Figure~\ref{fig:fes_bg}
the color distributions of all DSC
candidates with the color distributions of the expected
contaminant population from the custom control fields (scaled to the
area of one field).  We find that the DSCs are generally {\it redder}
than the background.  In most cases, the color distributions of
objects with $(g$--$z) < 1.0$  are well-matched by the colors of
objects in the control sample.  However, at redder colors, there is a
clear excess of objects.  This argues that the DSC candidates cannot
simply be a result of an underestimated background that has a higher
cosmic variance than expected.

%%%%%%%%%%%%%%%%%%%%%%%%%%%%%%%%%%%%%%%%%%%%%%%%%%%%%%%%%%%%%%%%%%%%%%%%%%
%%%%%%%%%%%%%%%%%%%%%%%%%%%%%%%%%%%%%%%%%%%%%%%%%%%%%%%%%%%%%%%%%%%%%%%%%%

%%%%%%%%%%%%%%%%%%%%%%%%%%%%%%%%%%%%%%%%%%%%%%%%%%%%%%%%%%%%%%%%%%%%%%%%%%
\begin{figure*}
%\epsscale{0.6}
\plottwo{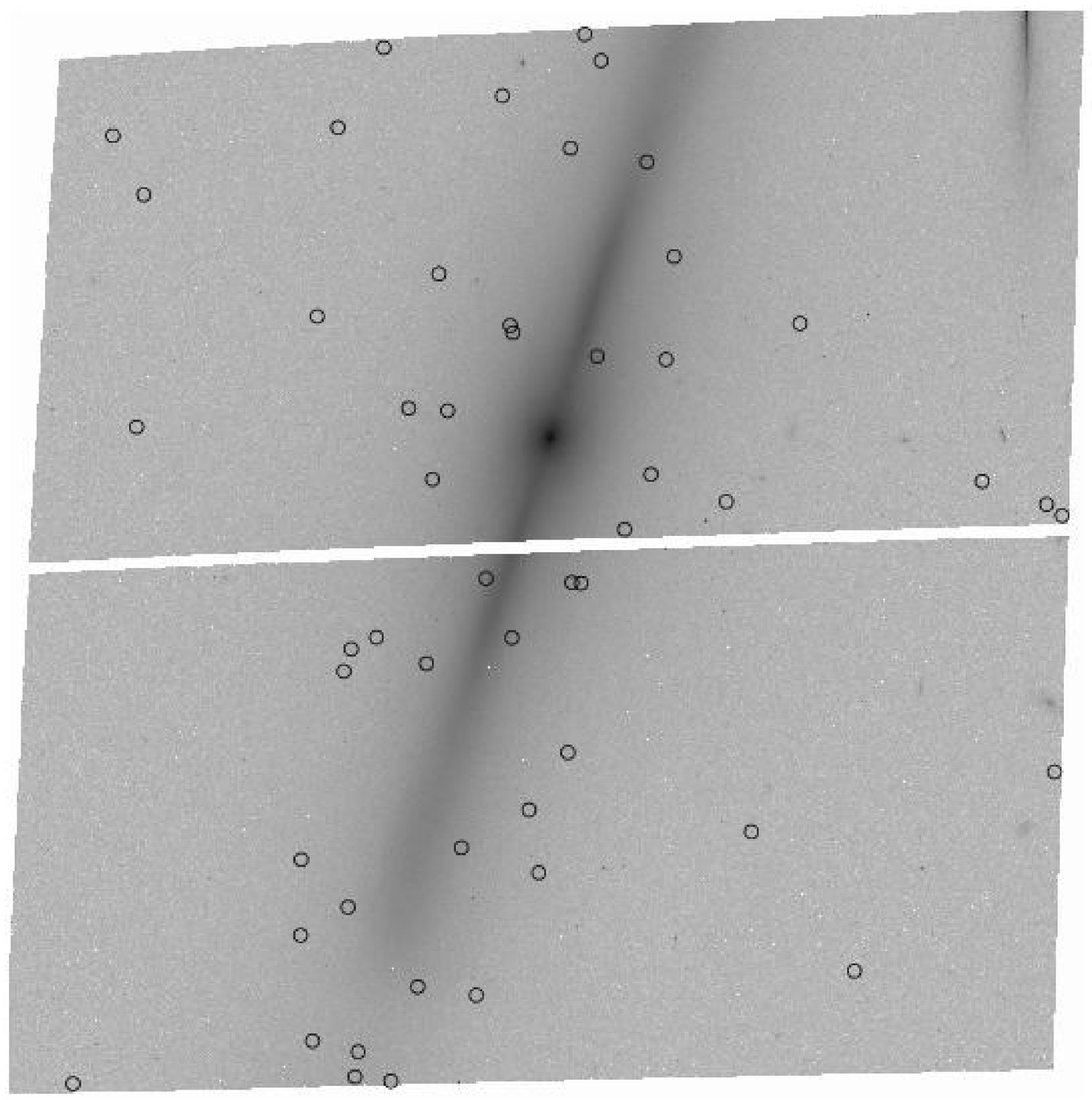}{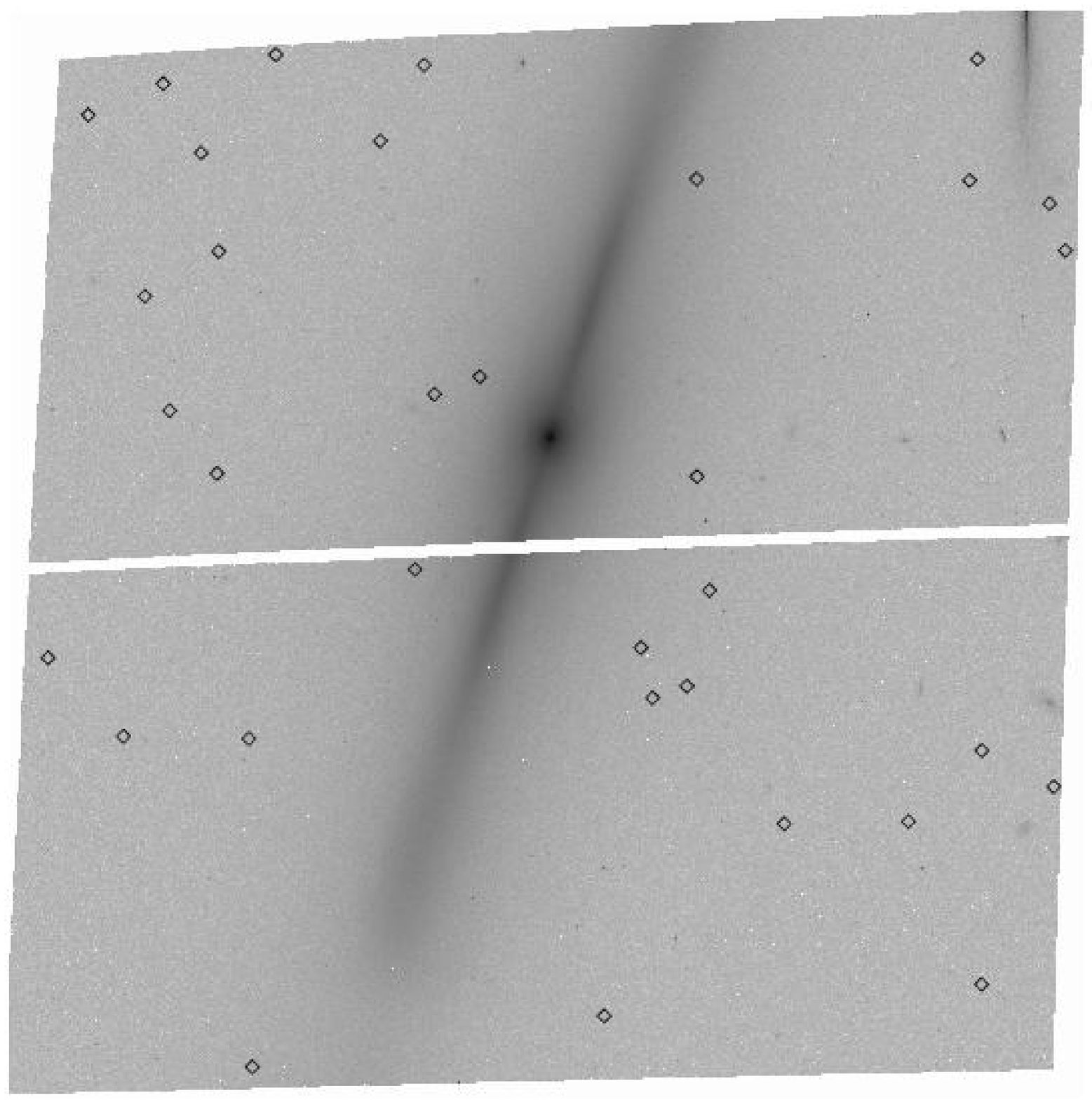}
\caption{Spatial distributions of diffuse sources around
  VCC~2095, one of the most edge-on galaxies in our sample.  Sources
  are divided by color at $(g$--$z)_0=1.0$.  Red sources
  ({\it left}) are generally in excess of the background, while blue sources 
  ({\it right}) are consistent with background levels seen in control
  fields.  The red sources are spatially clustered around the
  disk of the galaxy, supporting the idea that they are intrinsic to
  the galaxy.  Blue sources are scattered more randomly through the
  field. \label{fig:vcc2095}}
\end{figure*}
%%%%%%%%%%%%%%%%%%%%%%%%%%%%%%%%%%%%%%%%%%%%%%%%%%%%%%%%%%%%%%%%%%%%%%%%%%

A background cluster of galaxies, it may be argued, could also give
rise to an excess of red objects.  However, in
Figure~\ref{fig:vcc2095}, we show the spatial distribution of red and
blue DSC candidates (divided at $(g$--$z)=1.0$) 
around the highly inclined S0 galaxy, VCC~2095.  The red objects
appear highly clustered around the disk of the galaxy, despite the
lower detection efficiency in those regions (due to the galaxy
light).  The blue objects are distributed more randomly about the
field.  This argues that the red and diffuse objects are
in fact star clusters associated with the galaxy.  We discuss the
spatial distribution of the DSCs around the other galaxies in
Section~\ref{sec:spatial}.

\begin{figure}
\plotone{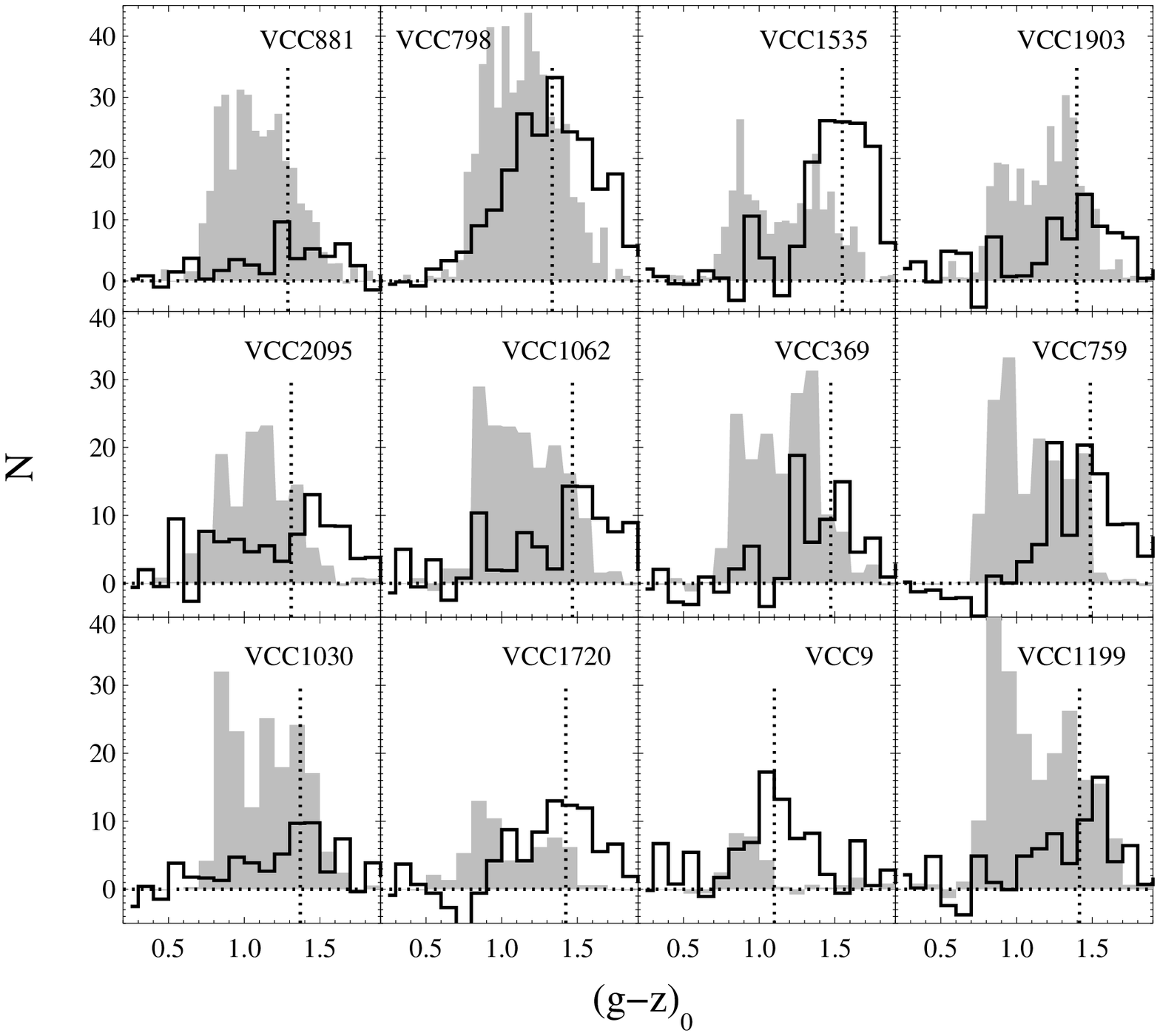}
\caption{Color distributions of diffuse star clusters (black) compared to
  colors of globular clusters (filled gray) for the 12 galaxies in
  Table~\ref{table:excesstable}.  Galaxies are ordered by decreasing
  $B$ luminosity as in Figure~\ref{fig:fes_bg}.  All distributions have been
  subtracted for the expected contamination from foreground and
  background sources.  All histograms use 0.1~mag bins except for the
  GCs for the four brightest galaxies (top row), which are binned in
  0.05~mag intervals.  Vertical dashed lines mark the median color of
  the diffuse sources.  This population is typically redder
  than even the red GC subpopulation.  \label{fig:fec_colors}}
\end{figure}

How do the colors of the DSCs compare to the other star clusters in
these galaxies, the globular clusters?  Once we take into account the
expected contamination from background galaxies, we can derive a
statistically corrected color distribution.
Figure~\ref{fig:fec_colors} shows the corrected DSC color
distributions for all twelve galaxies, and the median color for each.
Also shown are the GC color distributions for the same galaxies.  In
most galaxies, it is possible to see the bimodality in the GC color
distribution (Paper~IX).  In all cases, the median color of the DSCs
is as red or redder than the red GC subpopulation. These median color
values are listed in Table~\ref{table:excesstable}.  All of these
values are redder than the background objects in our control
fields, which have median colors of $(g$--$z) = 1.0$.  

%%%%%%%%%%%%%%%%%%%%%%%%%%%%%%%%%%%%%%%%%%%%%%%%%%%%%%%%%%%%%%%%%%%%%%%%%%
%%%%%%%%%%%%%%%%%%%%%%%%%%%%%%%%%%%%%%%%%%%%%%%%%%%%%%%%%%%%%%%%%%%%%%%%%%

\begin{figure}
\epsscale{1.2}
\plotone{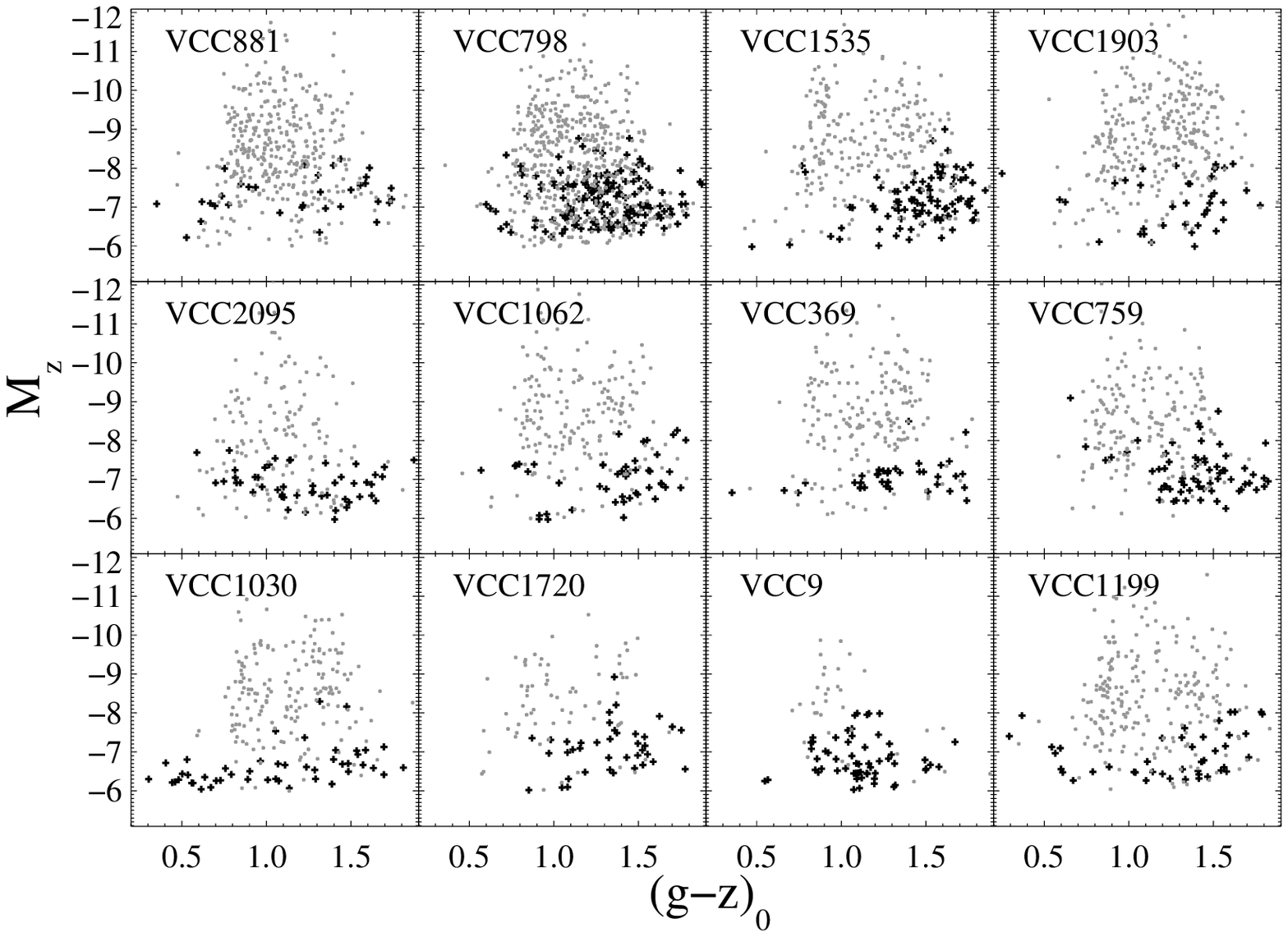}
\caption{Color--magnitude diagram, $(g$--$z)$--$M_z$, of 
  globular clusters (gray or red) and diffuse
  star clusters (black) in the twelve galaxies listed in
  Table~\ref{table:excesstable}.  Both samples have been statistically
  cleaned of background objects using control fields.  Globular
  clusters have $P_{GC}>0.5$, while the diffuse
  cluster population is defined to have $r_h > 4$~pc and $P_{GC} <
  0.2$.  The diffuse population is typically fainter than the turnover
  of the GC 
  luminosity function, and as red or redder than the red GCs.
  \label{fig:fec_cmd}}
\end{figure}

\begin{figure}
\epsscale{1.2}
\plotone{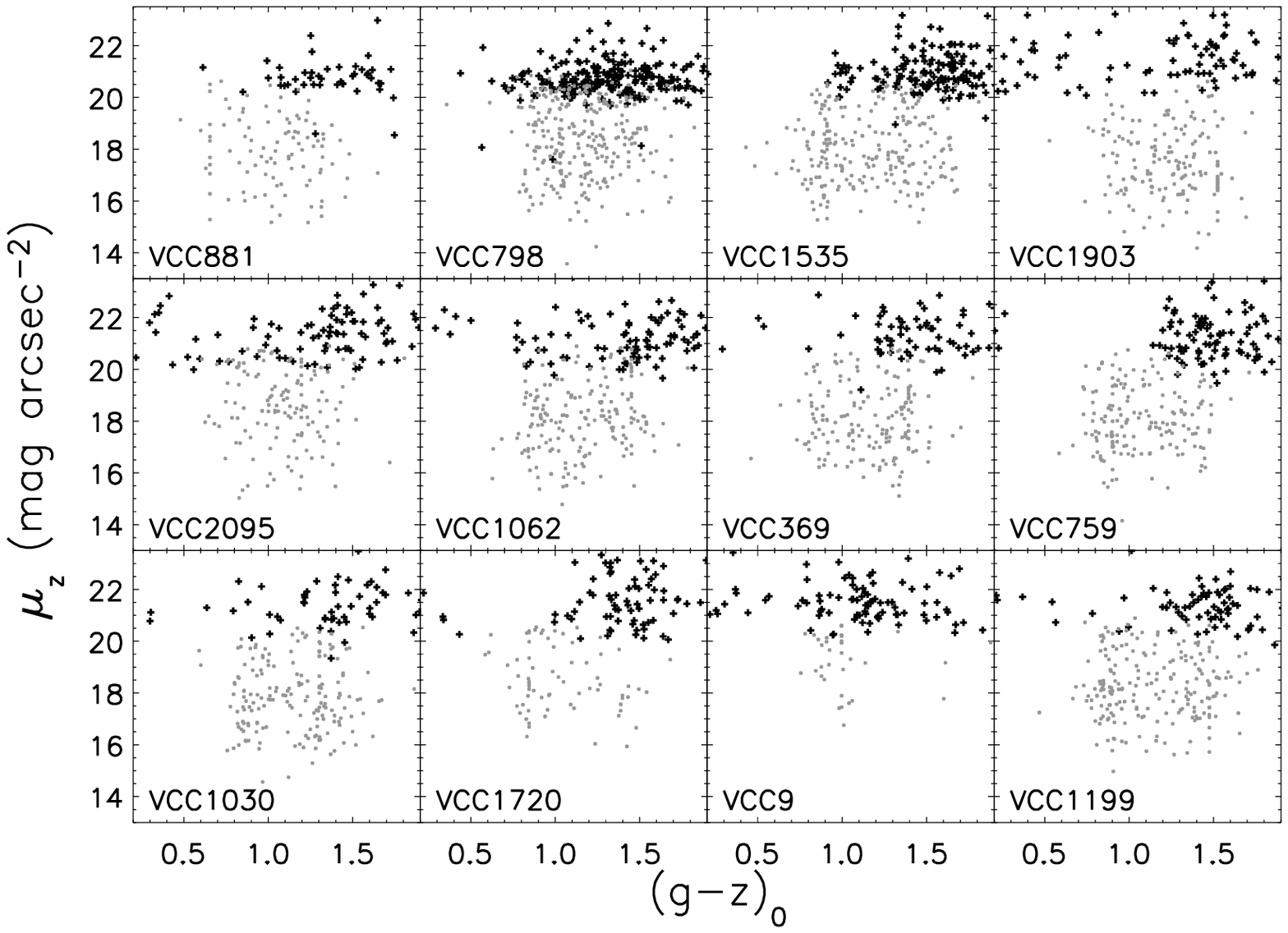}
\caption{Color--surface brightness diagram, $(g$--$z)_0$ versus $\mu_z$, of 
  globular clusters (red) and diffuse star clusters (black) 
  in the twelve galaxies listed in
  Table~\ref{table:excesstable}.  Data are the same as for
  Figures~\ref{fig:selectionfig}~and~\ref{fig:fec_cmd}. 
  The diffuse clusters often have very red colors, although some
  galaxies have diffuse clusters with a wide range of colors.  
  \label{fig:fec_colsb}}
\end{figure}

These properties are also displayed in Figures~\ref{fig:fec_cmd} and
\ref{fig:fec_colsb} where we show the color-magnitude and color-surface
brightness 
relationships for the star clusters in all twelve galaxies.  In both
figures, we compare the DSC and GC populations.  All plots have been 
statistically cleaned for background contamination using a ``nearest
neighbor'' approach --- for each object in a randomly selected subset
of the control catalog (one in seventeen for the 17 control fields),
we remove the object in our source catalogs that is closest in the
parameter space.  In Figure~\ref{fig:fec_cmd}, we see that the DSCs
are fainter and redder than the GCs.
Figure~\ref{fig:fec_colsb} shows color against $\mu_z$,
and the diffuse star clusters again generally appear
redder, although this is not universally true for all galaxies.  All
twelve galaxies have red DSCs, but some also have a number of blue
objects (in the case of VCC~1903, some are very blue).  At these
colors, contamination and Poisson noise 
from the background is more severe, so it
is difficult to draw conclusions about the nature of these objects.

%%%%%%%%%%%%%%%%%%%%%%%%%%%%%%%%%%%%%%%%%%%%%%%%%%%%%%%%%%%%%%%%%%%%%%%%%%
%%%%%%%%%%%%%%%%%%%%%%%%%%%%%%%%%%%%%%%%%%%%%%%%%%%%%%%%%%%%%%%%%%%%%%%%%%

\subsection{Spatial Distributions of DSCs}
\label{sec:spatial}

%%%%%%%%%%%%%%%%%%%%%%%%%%%%%%%%%%%%%%%%%%%%%%%%%%%%%%%%%%%%%%%%%%%%%%%%%%

%%%%%%%%%%%%%%%%%%%%%%%%%%%%%%%%%%%%%%%%%%%%%%%%%%%%%%%%%%%%%%%%%%%%%%%%%%

%%%%%%%%%%%%%%%%%%%%%%%%%%%%%%%%%%%%%%%%%%%%%%%%%%%%%%%%%%%%%%%%%%%%%%%%%%
\begin{figure}
\epsscale{1.2}
\plotone{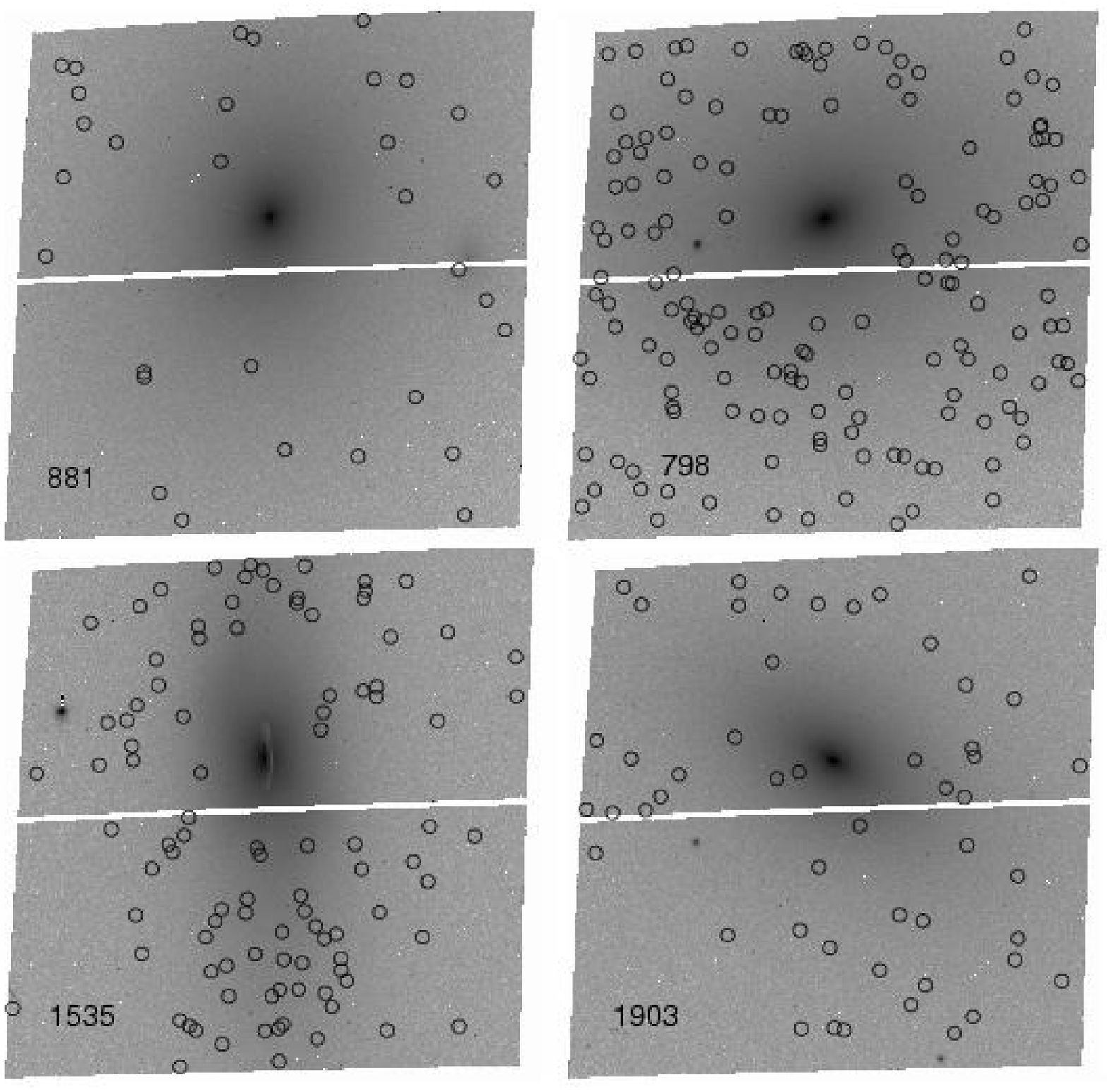}
\caption{Spatial distribution of red ($g$--$z > 1.0$), diffuse
  cluster candidates around galaxies with significant excesses.  While
  some of the objects are in the background, the large majority are
  likely star clusters.  For obviously inclined or elliptical
  galaxies, it is possible to see that the spatial distribution of the
  clusters appears to follow
  the underlying galaxy light.  The VCC numbers of the galaxies are
  shown in the lower left. \label{fig:spatialpic}}
\end{figure}
%%%%%%%%%%%%%%%%%%%%%%%%%%%%%%%%%%%%%%%%%%%%%%%%%%%%%%%%%%%%%%%%%%%%%%%%%%

%%%%%%%%%%%%%%%%%%%%%%%%%%%%%%%%%%%%%%%%%%%%%%%%%%%%%%%%%%%%%%%%%%%%%%%%%%
\begin{figure}
\epsscale{1.2}
\plotone{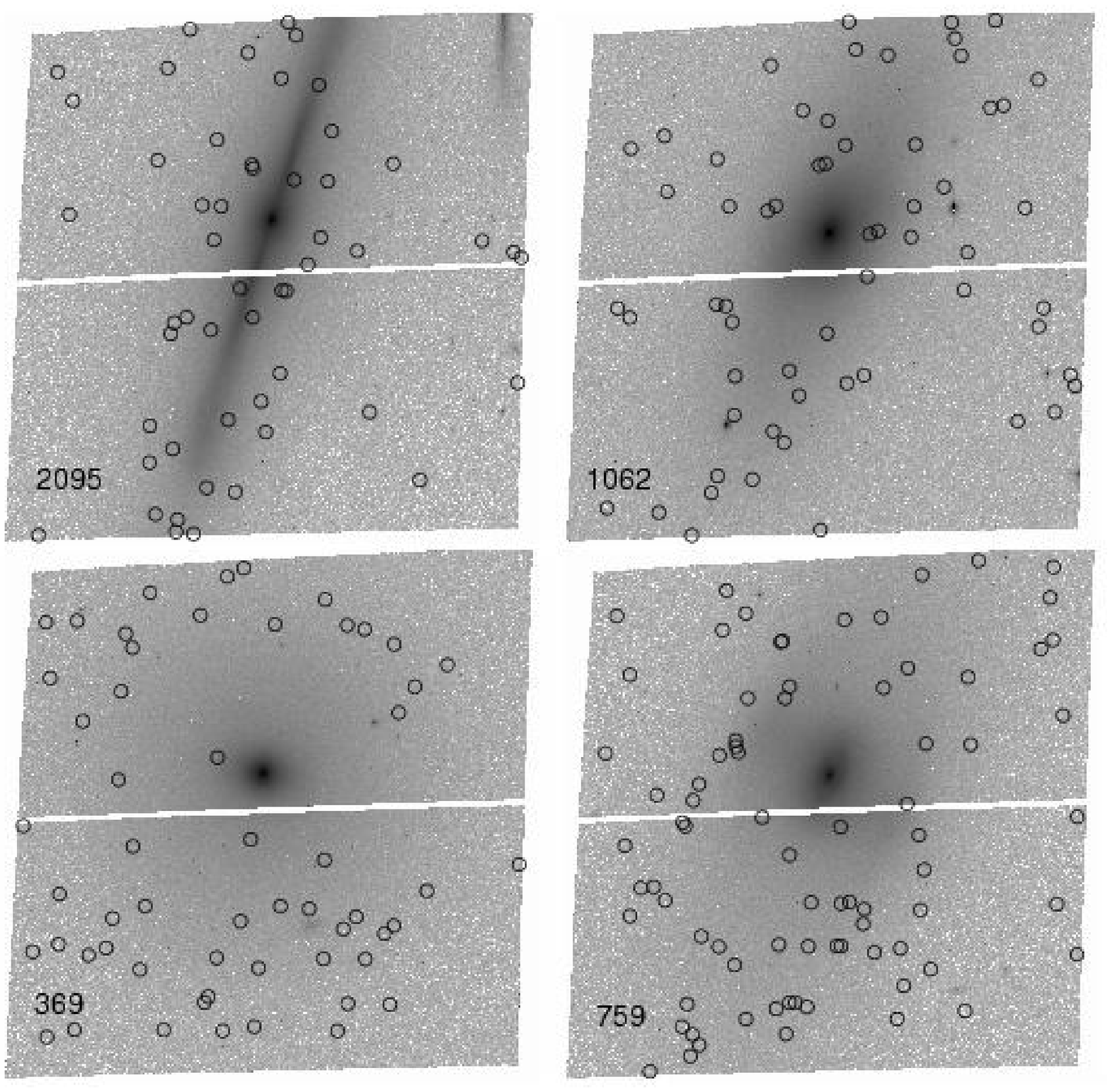}
\figurenum{11}
\caption{continued.  Spatial distribution of red ($g$--$z > 1.0$), diffuse
  cluster candidates around galaxies with significant excesses.}
\end{figure}

\begin{figure}
\epsscale{1.2}
\plotone{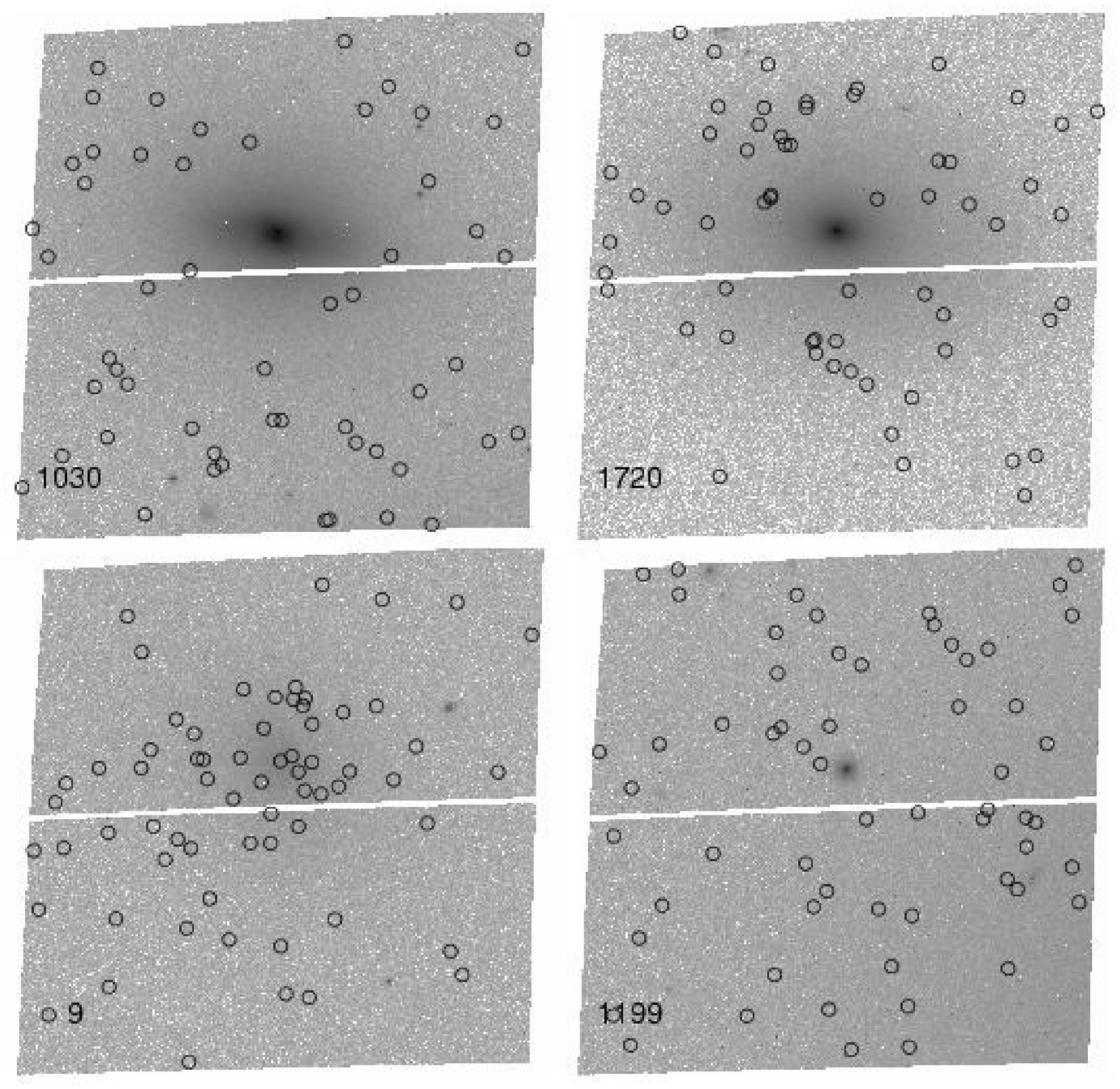}
\figurenum{11}
\caption{continued.  Spatial distribution of red ($g$--$z > 1.0$), diffuse
  cluster candidates around galaxies with significant excesses.}
\end{figure}
%%%%%%%%%%%%%%%%%%%%%%%%%%%%%%%%%%%%%%%%%%%%%%%%%%%%%%%%%%%%%%%%%%%%%%%%%%

By selecting only red DSC candidates ($g$--$z >1.0$), 
we can decrease the contaminant
fraction and investigate the spatial distribution of the DSCs.
Figure~\ref{fig:spatialpic} shows the positions of the red DSCs
plotted on the ACS/WFC images of the twelve ``excess'' galaxies.  In
galaxies that are significantly inclined or where the light
distribution is anisotropic, we find that the DSCs also have an
anisotropic distribution.  This particularly appears to be the case
for VCC~1535, 2095, 1062, and 1720.  Alignment with the galaxy
light in S0s suggests that the DSCs are associated with the stellar
disks of the galaxies, rather than the halos.  We note that when we
restrict ourselves to only red DSC candidates, two more galaxies meet
our criteria of having a $3\sigma$ excess over the mean---VCC~1154 and
VCC~2092, both of which are classified as S0s, and one of which (1154)
harbors a large kiloparsec-scale central dust disk.

Previous work on ``faint fuzzy'' clusters in NGC~1023 and 3384
(Larsen \& Brodie 2000; Brodie \& Larsen 2002) 
suggested that they were not only associated with the stellar disks of
these galaxies, but that they resided in a {\it ring} around the
galaxy, avoiding the central regions.  This has many
astrophysical implications if true (Burkert, Brodie, \& Larsen 2005)
so we wished to test this for the diffuse star clusters in the ACSVCS data.  

Because the central regions of galaxies have the highest surface
brightnesses, it is natural that we should detect fewer objects
there.  We estimate the detection efficiency in our images by adding
artificial star clusters at a range of magnitudes, half-light radii,
and background surface brightnesses.  We then run our detection
algorithm on the images and record the fraction that are recovered.
We are thus able to characterize the detection probability in this
three-dimensional parameter space.  

\begin{figure}
\plotone{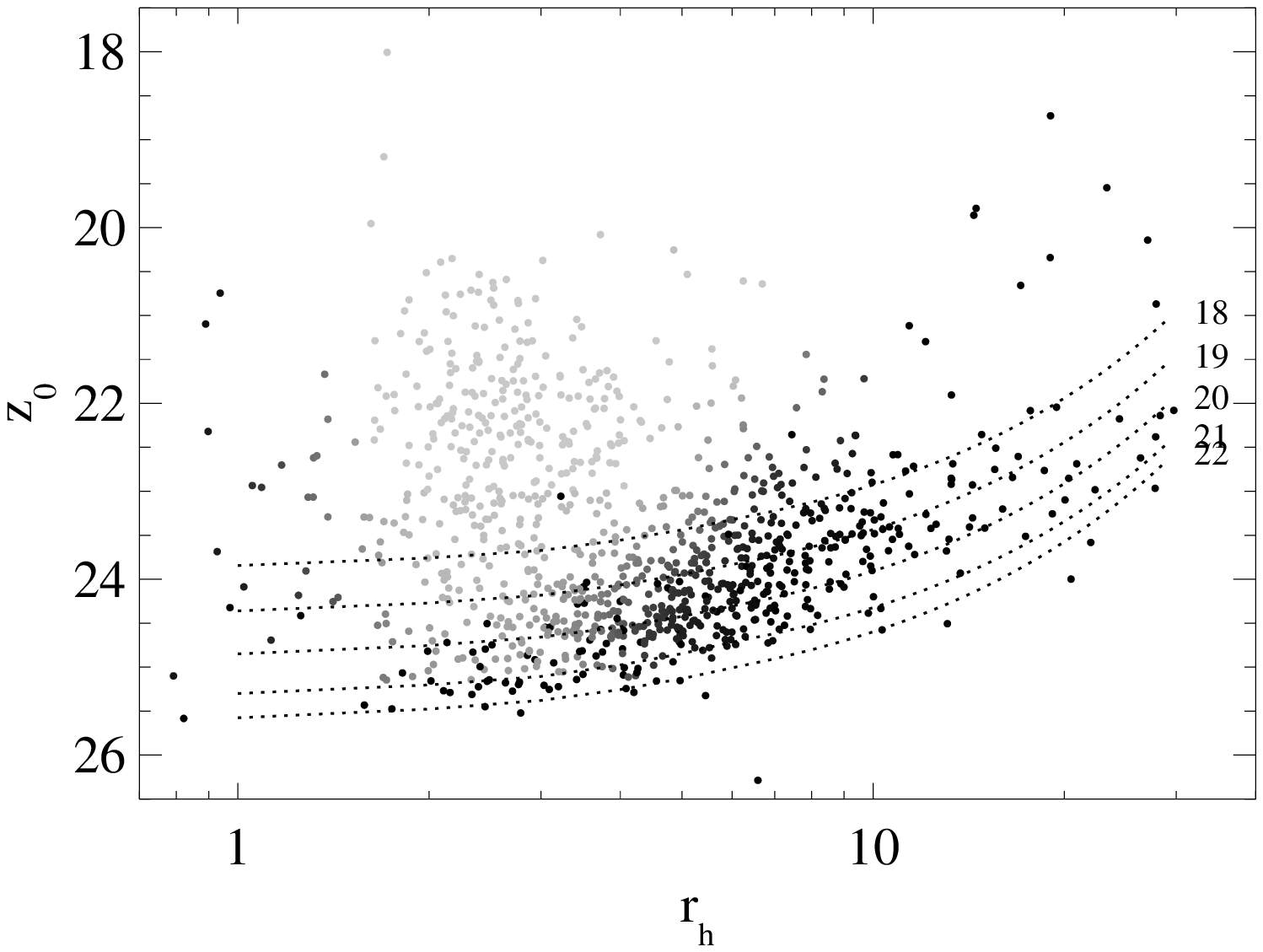}
\caption{Size-magnitude diagram for VCC~798 with 90\% completeness
  curves for different galaxy surface brightnesses.  This plot is the
  same as the bottom left of Figure~\ref{fig:m49m85} except that we
  have now overplotted our 90\% completeness curves for underlying
  galaxy surface brightnesses of 
  $\mu_{z,galaxy} = 18,\, 19,\, 20,\, 21,\, 
  {\rm and}\ 22\ {\rm mag\ arcsec}^{-2}$ as labeled.
  Notice that at $\mu_{z,galaxy}=19$, most of the diffuse clusters
  become difficult to detect.  In Figure~\ref{fig:iso90}, we show where
  this isophotal surface brightness level lies in the galaxy.
  \label{fig:sbdemo}}
\end{figure}

\begin{figure}
\plotone{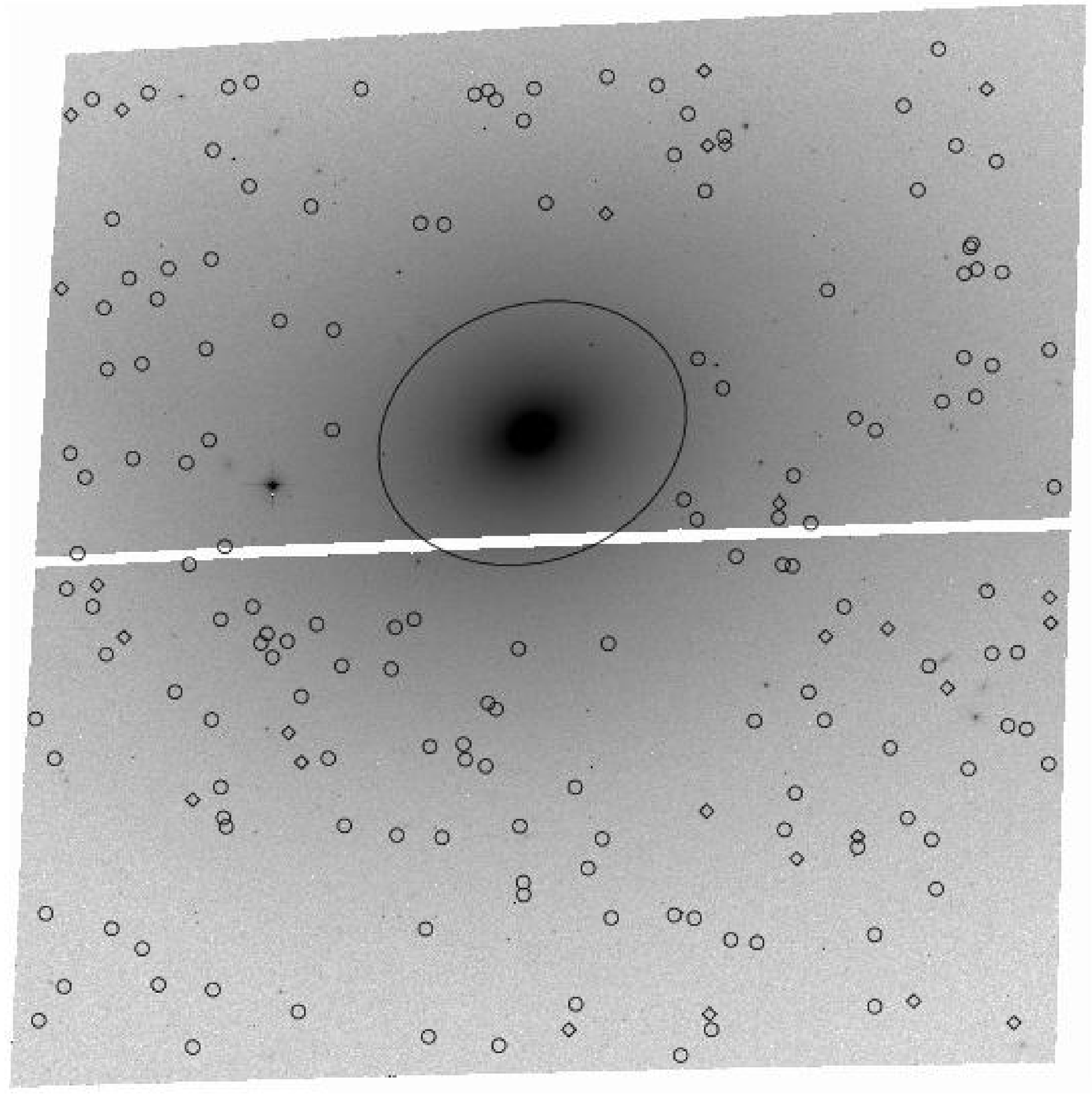}
\caption{F850LP ($z$) image of VCC~798 with diffuse objects
    and a $\mu_{z,galaxy}=19\ {\rm mag\
    arcsec}^{-2}$ isophote overplotted.  Inside this ellipse, we do not
    find any diffuse clusters, but this is likely because our
    detection efficiency is rapidly becoming low, and not necessarily 
    because of any intrinsic deficit of clusters toward the galaxy center.
    \label{fig:iso90}}
\end{figure}

In Figure~\ref{fig:sbdemo}, we show the size-magnitude diagram for
VCC~798, the galaxy with the largest number of DSCs.  We overplot the
90\% completeness curves in this plane for a range of background
surface brightnesses.  In the bright central regions of the galaxy
with $\mu_{z,galaxy}<19\ {\rm mag\ arcsec}^{-2}$, although over half of the
GCs are still detected, almost 
all of the DSC candidates fall below the detection threshold.  
Figure~\ref{fig:iso90} shows the ACS/WFC F850LP image of VCC~798 with the
locations of the DSC candidates.  The central ellipse corresponds to
the location of the $\mu_{z,galaxy}=19\ {\rm mag\ arcsec}^{-2}$ isophote using data
from Ferrarese \etal (2005).  The deficit of DSCs within this
isophote is due to incompleteness.  Thus, our data are not deep enough
to determine whether DSCs exist in the central regions of VCC~798 and
other luminous galaxies.  We do have one lower surface brightness
galaxy with a large number of DSCs, VCC~9, and
Figure~\ref{fig:spatialpic} shows that we
detect numerous DSCs clustered around that galaxy's center.  VCC~9, however,
appears different from the other galaxies in a number of ways,
however, so its properties may not be extendable to the rest of the sample.

%%%%%%%%%%%%%%%%%%%%%%%%%%%%%%%%%%%%%%%%%%%%%%%%%%%%%%%%%%%%%%%%%%%%%%%%%%
%%%%%%%%%%%%%%%%%%%%%%%%%%%%%%%%%%%%%%%%%%%%%%%%%%%%%%%%%%%%%%%%%%%%%%%%%%

%%%%%%%%%%%%%%%%%%%%%%%%%%%%%%%%%%%%%%%%%%%%%%%%%%%%%%%%%%%%%%%%%%%%%%%%%%
%%%%%%%%%%%%%%%%%%%%%%%%%%%%%%%%%%%%%%%%%%%%%%%%%%%%%%%%%%%%%%%%%%%%%%%%%%

\subsection{Specific Frequencies and Luminosities of DSCs}

The number of star clusters usually scales with the total luminosity of
the galaxy.  For globular clusters, it is sometimes useful to talk in
terms of specific frequency, a quantity scaled to unity for 1~GC in a
galaxy of $M_V=-15$.  Likewise, we would like to compare not just total
numbers of diffuse star clusters, but the number per unit galaxy
luminosity in order to investigate the efficiency of DSC formation.  In
Table~\ref{table:excesstable}, we list an ``observed'' specific frequency,
$S_{N,DSC}$, which is simply $N_{DSC}\times10^{0.4(M_V+15)}$, where
$N_{DSC}$ is the number of diffuse star cluster candidates that we
observe in excess of the background.  The galaxy magnitudes that we use
are the total measured $g$ and $z$ band fluxes observed {\it within the
ACS/WFC aperture}.  Because we do not observe the entire galaxy, the best
we can do is compute specific frequency within the ACS/WFC field of view.

For GCs, the luminosity function is well-approximated by a Gaussian, so it
is meaningful to talk about total numbers of GCs.  In the case of the
DSCs, we do not know the form of their luminosity function, so the
number that we observe is highly dependent on the depth of the
observation.  For both GCs and young massive
clusters in spiral galaxies, though, most of the luminosity is in the brightest
objects (i.e.\ the bright ends of their luminosity functions
follow a power-law with
exponents $>-2$).  This is why the specific luminosity of a star cluster
system is often a more robust quantity.  Following Larsen \& Richtler
(2000) and Harris (1991), we calculate the DSC specific luminosity in
the $z$-band,
which is proportional to ratio of the luminosity in diffuse clusters to
the galaxy's luminosity --- $T_{L,DSC}(z) = L_{z,DSC}/L_{z,galaxy}\times
10^4$.  We find that the fraction of galaxy light in DSCs for the nine
lenticular galaxies to be
$1$--$7\times 10^{-4}$.  When normalized to galaxy luminosity, the
galaxy with the highest number of DSCs, VCC~798, does not appear
particularly special with $T_{L,DSC}=4.2$.  The real outliers are VCC~9
and 1199.  This is to be expected for VCC~1199 if the star clusters we
see are really part of the M49 system.  VCC~9's very high specific
luminosity of DSCs suggests that it was either more efficient at
producing DSCs, or that it was once a more luminous system.

If specific luminosities of diffuse star cluster systems is relatively
constant across galaxies, then our 
$3\sigma$ criterion will make us insensitive to DSC systems in faint
galaxies where the noise is dominated by background galaxies.  For
instance, if the faintest galaxy in our sample (VCC~1661) had the same
specific frequency as VCC~1720 ($0.51$), then we would only expect it to have
0.6 diffuse clusters.  Thus, the dwarfs in our sample may have a few
DSCs, but it is nearly impossible to draw firm conclusions.

\subsection{Properties of Galaxies with Diffuse Star Clusters}
\subsubsection{Morphology}

%%%%%%%%%%%%%%%%%%%%%%%%%%%%%%%%%%%%%%%%%%%%%%%%%%%%%%%%%%%%%%%%%%%%%%%%%%
\begin{figure}
\epsscale{1.2}
\plotone{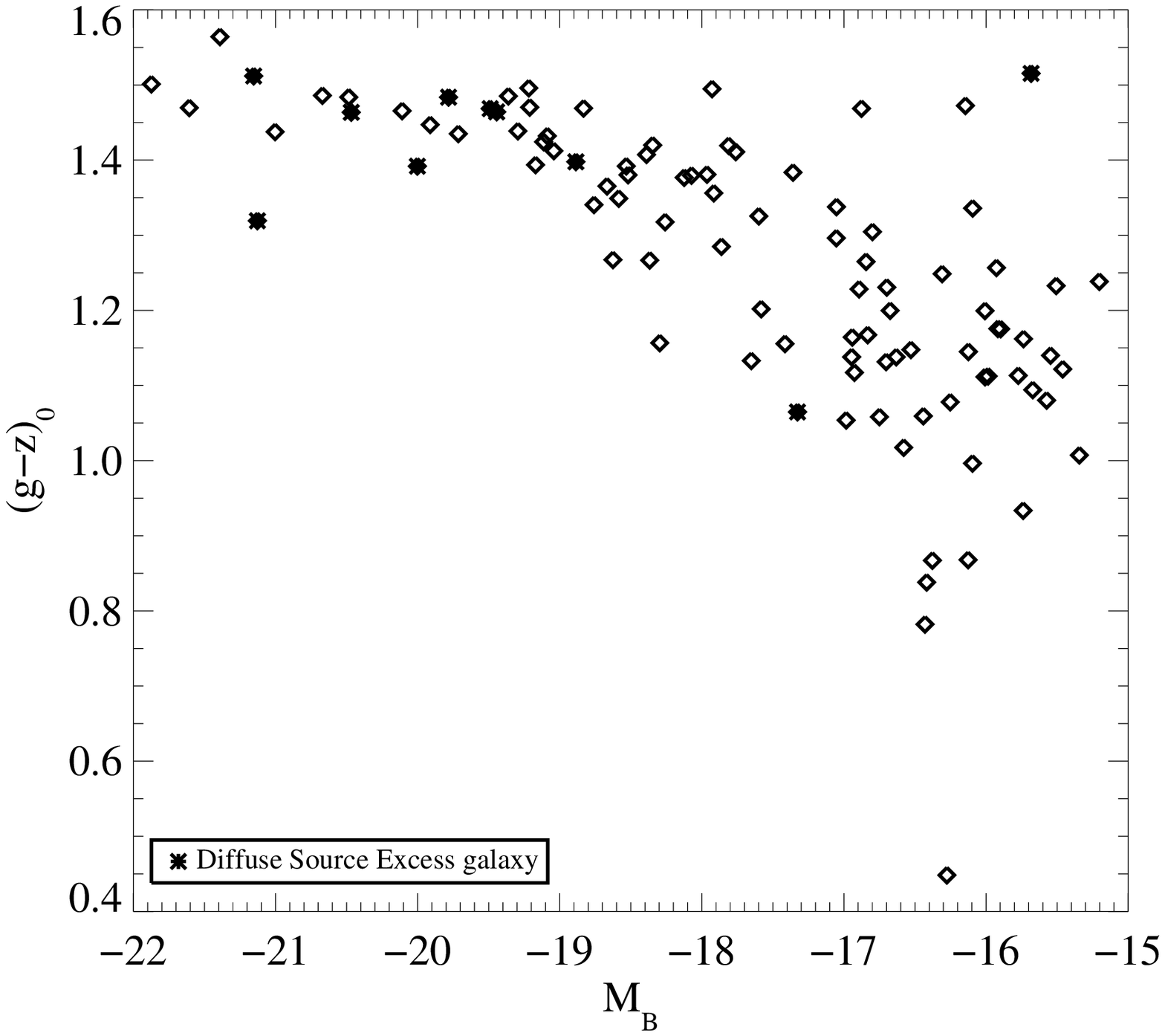}
\caption{Color-magnitude diagram of galaxies with an excess of
  diffuse sources, compared to the rest of the ACSVCS sample.  Colors
  are from Paper~VI and absolute B magnitudes are from BST85 assuming
  a distance of 16.5~Mpc.  While two of the galaxies appear to be
  significantly bluer than the mean color-magnitude relation (VCC~798
  and VCC~9), other galaxies with DSCs do not have colors that distinguish
  themselves from normal galaxies.  VCC~1199 in the upper left is
  a companion of M49 (VCC~1226) and is known to be much redder than
  expected for its luminosity.\label{fig:fec_galxcmd}}
\end{figure}
%%%%%%%%%%%%%%%%%%%%%%%%%%%%%%%%%%%%%%%%%%%%%%%%%%%%%%%%%%%%%%%%%%%%%%%%%%

To try to understand the origin of the DSC population, we looked at the
different properties of the host galaxies.  Most of the galaxies that
have significant numbers of diffuse star clusters 
are lenticular (S0) galaxies, pointing
to a possible connection to the existence of a stellar disk.  The
spatial clustering around the disks of inclined galaxies supports this
view.  However, there are many S0 galaxies in our sample that do 
{\it not} show significant diffuse star cluster populations.
Figure~\ref{fig:nvsmb} shows a grouping of galaxies with 
$-19.5< M_B<-19$, many of which have S0 morphological classifications, but
none of which have a very significant number of DSCs.  This
corroborates the findings of Larsen \etal
(2001) who also found that while the two lenticulars NGC~1023 and 3384 
had ``faint fuzzies'', another lenticular, NGC~3115, did not possess
them, implying that they are not a universal phenomenon in S0 galaxies.

To make a comparison between similar galaxies that have different DSC
properties, we create a
comparison sample of eight intermediate luminosity early-type galaxies with 
$-19.5< M_B<-19$ of which seven are classified as S0, E/S0 or S0/E
(Table~\ref{table:comptable}), and all of which have low numbers of DSC
candidates.  
The median excess above the mean level for faint, extended sources in
these galaxies is $0.9\sigma$. This is compared to a sample of
galaxies with DSCs that are in the magnitude range $-20.1 < M_B <
-18.8$ which have a median excess of $4.2\sigma$ above the mean value.
We limit the magnitude range of the two samples to make a meaningful
comparison between the globular cluster populations because we know
that the properties of the galaxies and their GC systems vary with
galaxy luminosity (e.g.\ Gebhardt \& Kissler-Patig 1999; Paper IX).  
Interestingly, only one of these galaxies, VCC~685,
contains any visible dust.

While morphology is a clue, it is not the full story.  In
Figure~\ref{fig:fec_galxcmd} we plot the color-magnitude diagram for
all ACSVCS galaxies and compare them to the galaxies with significant
numbers of DSCs.  If the existence of DSCs is related to recent
episodes of star formation, we might expect that these galaxies would
have bluer than average colors.  This appears to be the case for
VCC~798 and VCC~9, but the other galaxies do not show any significant
deviation from the color-magnitude relation of early-type galaxies.
VCC~1199 is a companion of VCC~1226 (M49) and is much redder than
expected for its luminosity.  

%%%%%%%%%%%%%%%%%%%%%%%%%%%%%%%%%%%%%%%%%%%%%%%%%%%%%%%%%%%%%%%%%%%%%%%%%%
\begin{deluxetable}{rrcc}
\tablewidth{0pt}
\tablecaption{Comparison Galaxies \label{table:comptable}}
\tablehead{
\colhead{VCC} & 
\colhead{$M_B$} & 
\colhead{$(g_{475}$--$z_{850})_0$} & 
\colhead{Type(VCC)} \\ 
\colhead{(1)} &
\colhead{(2)} &
\colhead{(3)} &
\colhead{(4)} 
}
\startdata
1692 & $-19.4$ &   1.51 & S0/E \\ 
2000 & $-19.3$ & \nodata & E/S0 \\ 
 685 & $-19.2$ &   1.46 & S0   \\ 
1664 & $-19.2$ &   1.39 & E    \\ 
 654 & $-19.2$ &   1.48 & S0   \\ 
 944 & $-19.1$ &   1.47 & S0   \\ 
1938 & $-19.1$ &   1.46 & S0   \\ 
1279 & $-19.0$ & \nodata & E    \\ 
\enddata
\tablenotetext{1}{Number in Virgo Cluster Catalog}
\tablenotetext{2}{Absolute B Magnitude, extinction-corrected, $D=16.5$~Mpc}
\tablenotetext{3}{Color from Paper VI}
\tablenotetext{4}{Morphological type from the VCC}
\tablenotetext{5}{Morphological type from the NED}
\end{deluxetable}

%%%%%%%%%%%%%%%%%%%%%%%%%%%%%%%%%%%%%%%%%%%%%%%%%%%%%%%%%%%%%%%%%%%%%%%%%%

\subsubsection{Globular Clusters}
%%%%%%%%%%%%%%%%%%%%%%%%%%%%%%%%%%%%%%%%%%%%%%%%%%%%%%%%%%%%%%%%%%%%%%%%%%
\begin{figure}
\epsscale{1.25}
\plotone{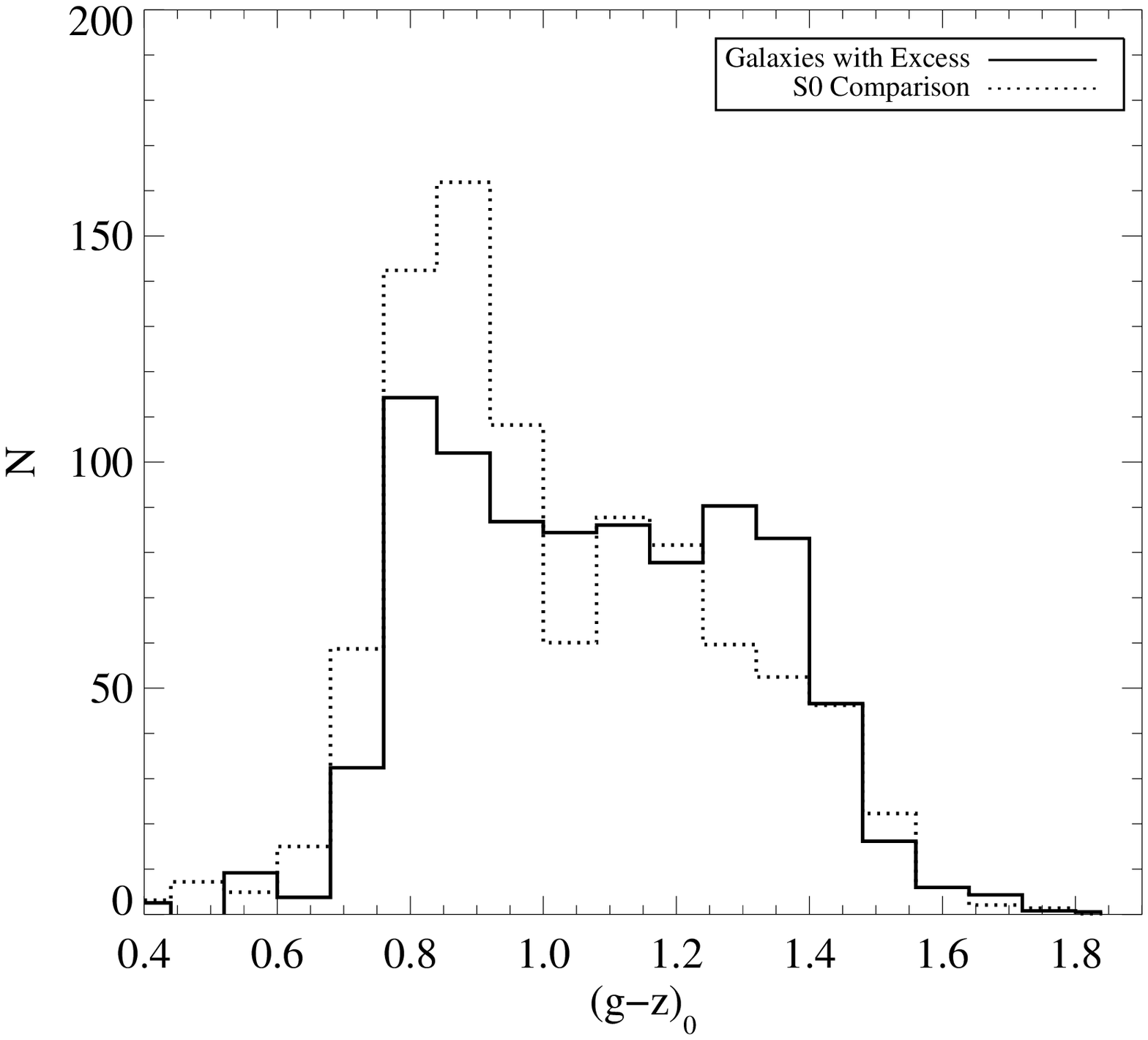}
\caption{Combined globular cluster color distributions for S0 galaxies
  with and without a significant population of diffuse
  clusters.  We choose only galaxies with $-21.1<M_B<-18.8$ to create
  a more uniform comparison.  The comparison sample of S0s is listed
  in Table~\ref{table:comptable}.  Both
  distributions appear bimodal and the means of the two peaks are very
  similar.  The galaxies with DSCs may have a slightly redder red peak,
  and an excess of red
  GCs, but this may also be due to the larger luminosity range of
  these galaxies, in particular the inclusion of VCC~1903 which has a
  significant red GC subpopulation.\label{fig:gccomp}}
\end{figure}
%%%%%%%%%%%%%%%%%%%%%%%%%%%%%%%%%%%%%%%%%%%%%%%%%%%%%%%%%%%%%%%%%%%%%%%%%%

Likewise, we might expect that the color distributions of globular
clusters in galaxies with and without an DSC population might be
substantially different.  
Figure~\ref{fig:gccomp} shows the
combined GC color distributions for the DSC and comparison galaxy
samples.  Both distributions are bimodal with the mean colors of the
blue peaks very similar.  The mean color of the red GCs appears
slightly redder in the DSC galaxy sample.  The most notable difference is that
the fraction of red GCs is lower in the comparison sample, a value of
0.45 as opposed to 0.56 for the galaxies with DSCs.  This, however,
may be a result of the larger luminosity range for the galaxies with
DSCs, particularly the inclusion of VCC~1903 which has a very high
fraction of red GCs (0.61) and contributes $\sim30\%$ ($269/911$) of
the GCs to that group.

\subsubsection{Environment}

Another factor that may play some role in the existence of
DSCs is the local environment.  Galaxies such as VCC~798 and 1030 have
nearby companion spiral galaxies.  VCC~798 is almost certainly
interacting with the neighboring SBb galaxy, NGC~4394, which lies
at a distance of $7\farcm6$ (37~kpc, projected).  VCC~798 also
exhibits fine structure that are indications of a recent interaction
(Schweizer \& Seitzer 1992).  VCC 1030 is $4\farcm4$ (21~kpc,
projected) from NGC~4438 (VCC~1043), an S0/a galaxy with a very
disturbed morphology.  
Other galaxies such as VCC~9, whose nearest neighbor in the VCC has a
projected distance of 81~kpc, are very isolated.

The first diagnostic of density in a galaxy cluster is the projected
distance to the cluster center.  For the Virgo Cluster, we take the
position of the cD galaxy M87 (VCC~1316) to be the center of the
cluster at $(\alpha,\delta)_{J2000}= (12$:$30$:$49.43,\ +12$:$23$:$28.35)$.
We do not detect any significant trend with clustercentric
radius, except perhaps for the fact that the galaxies with the two largest
excesses (VCC 798 and 1535) are far from the center of the cluster.

We also use the Virgo Cluster Catalog (BST85) to define a metric of
the local environmental density.  We use all galaxies in the VCC that
are certain or probable members of the cluster, and by determining the
distance to the first, fifth, and tenth nearest neighbor we can calculate the
local surface density of galaxies within that radius.
We do not find correlations between any of these density indicators and
the number of DSCs.  So, despite the clues given to us by
VCC~798 and 1030, it does not appear that local galaxy density is a
strong indicator of the existence of DSCs.

%%%%%%%%%%%%%%%%%%%%%%%%%%%%%%%%%%%%%%%%%%%%%%%%%%%%%%%%%%%%%%%%%%%%%%%%%%
\begin{figure}
\epsscale{1.25}
\plotone{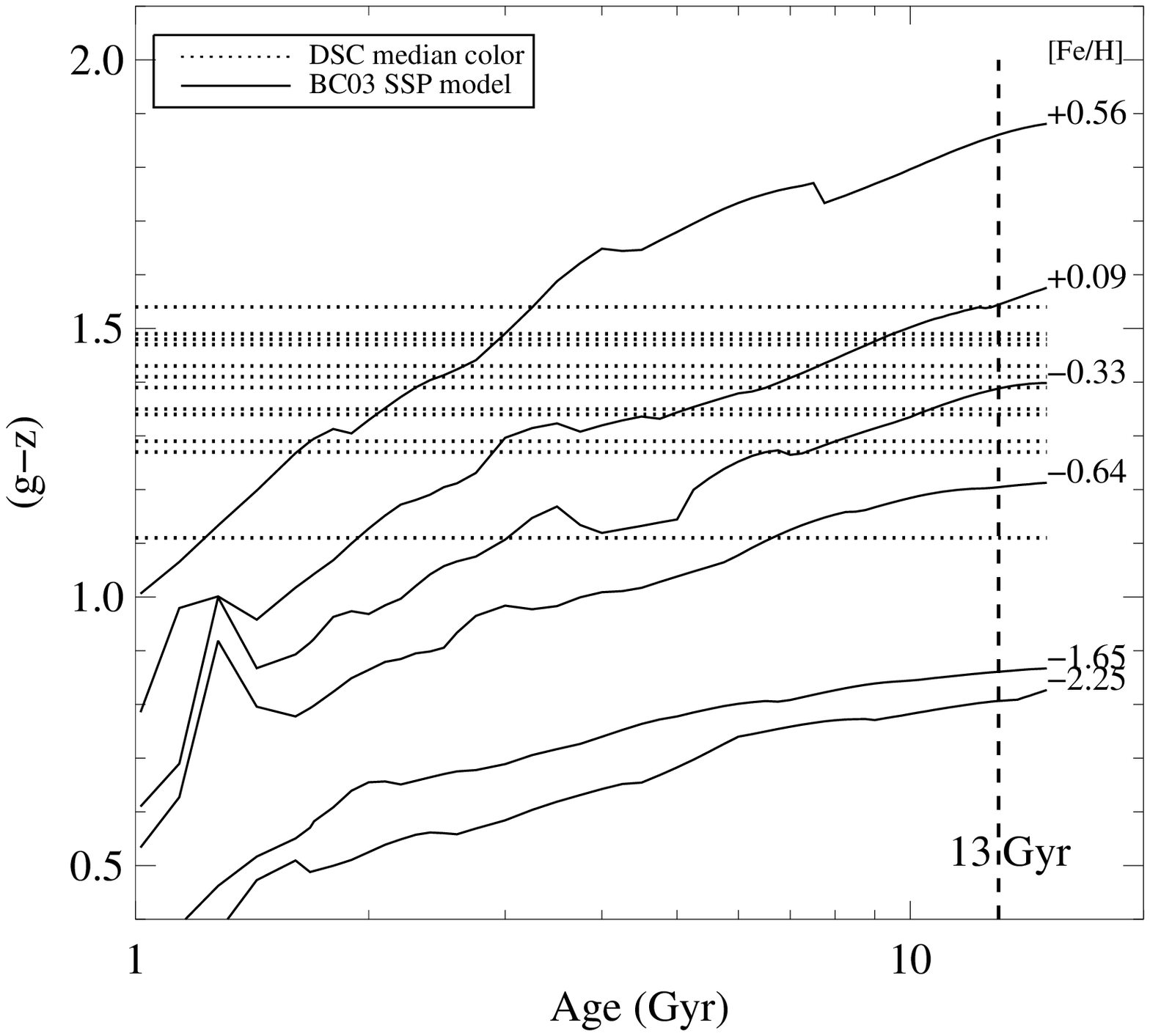}
\caption{Comparison of the colors of diffuse clusters with
  simple stellar population (SSP) models of Bruzual and Charlot (2003).  Solid
  lines represent isometallicity SSP tracks showing the evolution of
  $(g$--$z)$ with age from 1--15 Gyr.  The vertical dashed line marks an
  age of 13~Gyr.  Horizontal dotted lines mark the median colors of the
  DSC populations from Table~\ref{table:excesstable}.  With the exception
  of VCC~9, which has the bluest color, most DSC populations must either
  have old ages ($>5$~Gyr) or metallicities substantially in excess of solar.
  \label{fig:fec_bc03}}
\end{figure}
%%%%%%%%%%%%%%%%%%%%%%%%%%%%%%%%%%%%%%%%%%%%%%%%%%%%%%%%%%%%%%%%%%%%%%%%%%

\section{Discussion}
\subsection{The Ages and Metallicities of DSCs}
Like most star clusters, DSCs are likely to be
simple stellar populations (SSPs), so even though 
we only have a single color with which to constrain their properties, we
can compare them to models and still gain reasonable constraints on
their possible ages and metallicities.  In Figure~\ref{fig:fec_bc03}, we
plot isometallicity tracks for SSPs from the models of Bruzual \&
Charlot (2003) using a Chabrier (2003) initial mass function. 
We use SSP models because there are no good empirical measurements of
star clusters in $(g$--$z)$ that cover a sufficient range in age and
metallicity.  For five
different metallicities, the models show how 
the $(g$--$z)$ color of SSPs become redder as the population ages from 1
to 15~Gyr.  Overplotted are lines representing the median colors of the
DSC populations from Table~\ref{table:excesstable}.
With the exception of VCC~9 which has by far
the bluest DSC population, {\it all the DSC systems must either have old
  ($>5$~Gyr) ages or metallicities in excess of solar}.  Even for
super-solar metallicities, it is very unlikely that the DSCs have mean ages
younger than 2~Gyr.  Using a
Salpeter IMF makes the models redder by about 0.04~mag but does not
change the general conclusion.  For VCC~1535,
the reddest DSC system, even with an age of 13~Gyr the clusters must have 
${\rm [Fe/H]}=+0.09$.  We note that VCC~1535 does harbors a central dust
disk, and although the DSCs are well beyond the visible dust in our
images we cannot rule out the possibility that there might be some
reddening that is unaccounted for.  However, there is no systematic
color difference for DSCs in galaxies with and without visible dust.
Nevertheless, we caution that with only two filters we are vulnerable to
any internal extinction that may exist.

We also find that the colors of the DSCs appear to closely track the
colors of their parent galaxy.  For half of the sample, the median DSC
color is almost exactly the color of the galaxy, implying
that the two stellar populations coevolved.  While the
age-metallicity degeneracy prevents us from drawing definitive
conclusions, if we make the assumption that the ages of the clusters are
between 8 and 13~Gyr, then their metallicities span a range of
approximately $-0.3 < {\rm [Fe/H]} < +0.1$.  The only exception is VCC~9,
whose DSCs must have ${\rm [Fe/H]}\gtrsim -0.8$, but could also have
ages of $\sim2$~Gyr at solar metallicity.

\subsection{Comparisons with the Milky Way}

%%%%%%%%%%%%%%%%%%%%%%%%%%%%%%%%%%%%%%%%%%%%%%%%%%%%%%%%%%%%%%%%%%%%%%%%%%
\begin{figure}
\epsscale{1.25}
\plotone{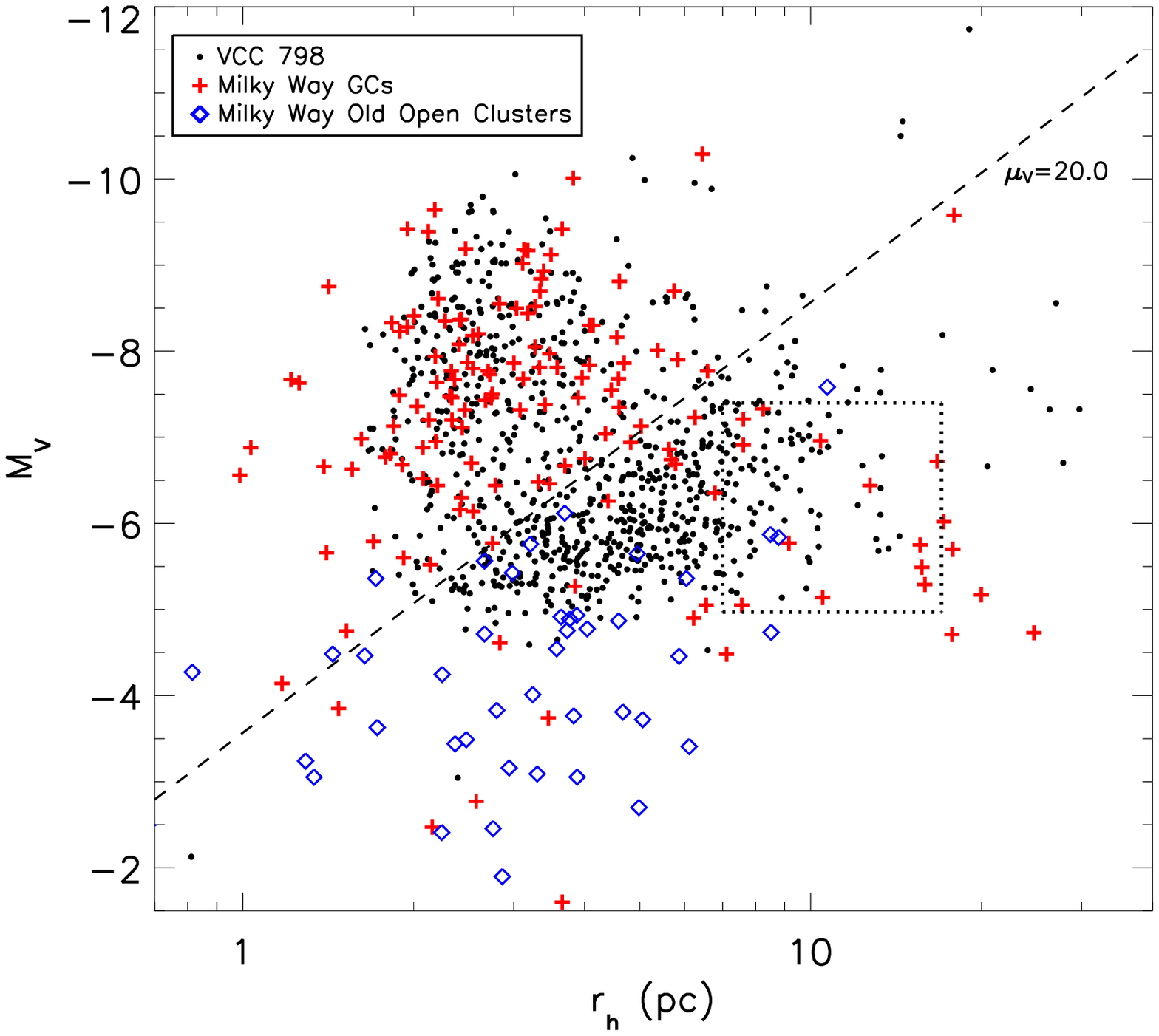}
\caption{Size-magnitude diagram for star clusters in
  VCC~798 (dot), Milky Way globular clusters (cross), Milky Way old open
  clusters (diamond), and ``faint fuzzy'' selection box from Larsen \& Brodie
  (dotted lines).
  The VCC~798 catalog was statistically cleaned of background
  contamination, and photometry was transformed
  to $M_V$ using the relations in the text and a distance of 16.5~Mpc.  
  The dashed line represents a constant mean
  surface brightness, $\mu_V = 20\ {\rm mag\ arcsec}^{-2}$, below which
  most DSCs lie.
  The locus of GCs in VCC~798 match the GCs in the Milky Way with the
  exception of the low surface brightness Galactic halo GCs which
  would fall below our detection threshold (lower right).  
  Most of the Milky Way old open clusters would also fall below our
  detection threshold.  The brightest of 
  these clusters, however, would have the sizes and magnitudes of DSCs
  in our sample.  In the Galaxy, there is a
  deficit of objects in the DSC locus.  The ``faint
  fuzzy'' region selects the more extended DSCs.
  \label{fig:mw_comp}}
\end{figure}
%%%%%%%%%%%%%%%%%%%%%%%%%%%%%%%%%%%%%%%%%%%%%%%%%%%%%%%%%%%%%%%%%%%%%%%%%%

One frequently raised question in the study of these diffuse star clusters
is whether they are truly different from the star clusters found in the
Milky Way.  Some of the
fainter globular clusters are quite extended, with half-light radii of
$7<r_h<30$~pc, and outer halo GCs are known to span a large range in surface
brightness  (e.g.\ van~den~Bergh \& Mackey 2004). 
These GCs, though, are associated with the halo and are
much more metal-poor than the DSCs appear to be.  
Open clusters are found in the Galactic disk and are
fainter than globular clusters so provide a tempting parallel.  Most
open clusters are unlike DSCs, though, in that they are 
only a few hundred Myr old.  The old open clusters, with ages of
$0.7$--$12$~Gyr (Janes \& Phelps 1994), 
might be more similar.  The combination of
the diffuse star clusters being old, 
metal-rich, {\it and} low surface brightness makes
them different from most known Milky Way analogs.  In Figure~\ref{fig:mw_comp},
we show the size-magnitude diagram for all star cluster 
candidates (GC and DSC) in VCC~798, the system in our sample
with the largest number of DSCs.
In order to facilitate comparisons with other published data, we
transform our ACS $g$ and $z$ magnitudes to Johnson $V$.  We derive this
transformation using the $V$ and $I$ photometry of M87 globular clusters
presented in Kundu \etal (1999) and our $g$ and $z$ photometry from the
ACSVCS.  We do a robust linear fit to the data and obtain
\bigskip

\begin{equation}
V = g + 0.026(\pm0.015) - 0.307(\pm0.013)\times(g-z)
\end{equation}
\begin{equation}
I = z - 0.458(\pm0.017) + 0.172(\pm0.014)\times(g-z).
\end{equation}

Figure~\ref{fig:mw_comp} also shows the locations in this diagram
of the Milky Way globular
clusters with data taken from the compilation of Harris (1996).  Most of
the Galactic GCs have $r_h$ around 3~pc and share this region of the
diagram with a large number of VCC~798 GCs.  The rest of the Galactic
GCs, most of which have $V>-6.5$ and $r_h>5$~pc, have very low surface
brightnesses and a large number are below our detection threshold.  The
locus of the VCC~798 DSCs, a band from (3~pc,$-5$~mag) to (30~pc,$-8$~mag), is
comparatively devoid of Galactic globular clusters.

We also show the locations of Milky Way old open
clusters in this diagram.  
We use diameters, distances, and reddenings from van den Bergh (2005b),
which is based on the catalogs of Dias \etal (2002).  We used $V$
photometry from the SIMBAD astronomical reference database.  Of the 74
old open clusters listed in Friel (1995), we were able to obtain these
data for 44 objects.  Ages for these clusters range from 0.8-8~Gyr.
Size estimates should be viewed with caution as they
are not true half-light radii, but are usually taken to be the radius
at which the cluster stellar density matches that of the field.  This
radius, though, is probably most associated with a half-light radius
as opposed to a core or tidal radius (Friel 1995).  
We caution that the data for this sample are very heterogeneous, but they are
sufficient to show that the bulk of the old open clusters are much
fainter than the DSCs we are seeing in the ACSVCS.  It is also likely
that the old open clusters are younger than the DSCs and will fade
even more in this diagram.  However, the surface brightness distribution
of the old open clusters does overlap with that of the DSCs, and 
it is plausible that the DSCs are
in fact just the most luminous objects of a much more massive system
of disk clusters that is akin to the old open clusters of the Milky Way.
Both deeper observations of external galaxies and a more homogeneous
and precise catalog of Galactic open cluster properties will be
needed to explore any further connection between DSCs and open clusters.

\subsection{Comparisons with Star Clusters in Other Galaxies}

\begin{figure}
\epsscale{1.25}
\plotone{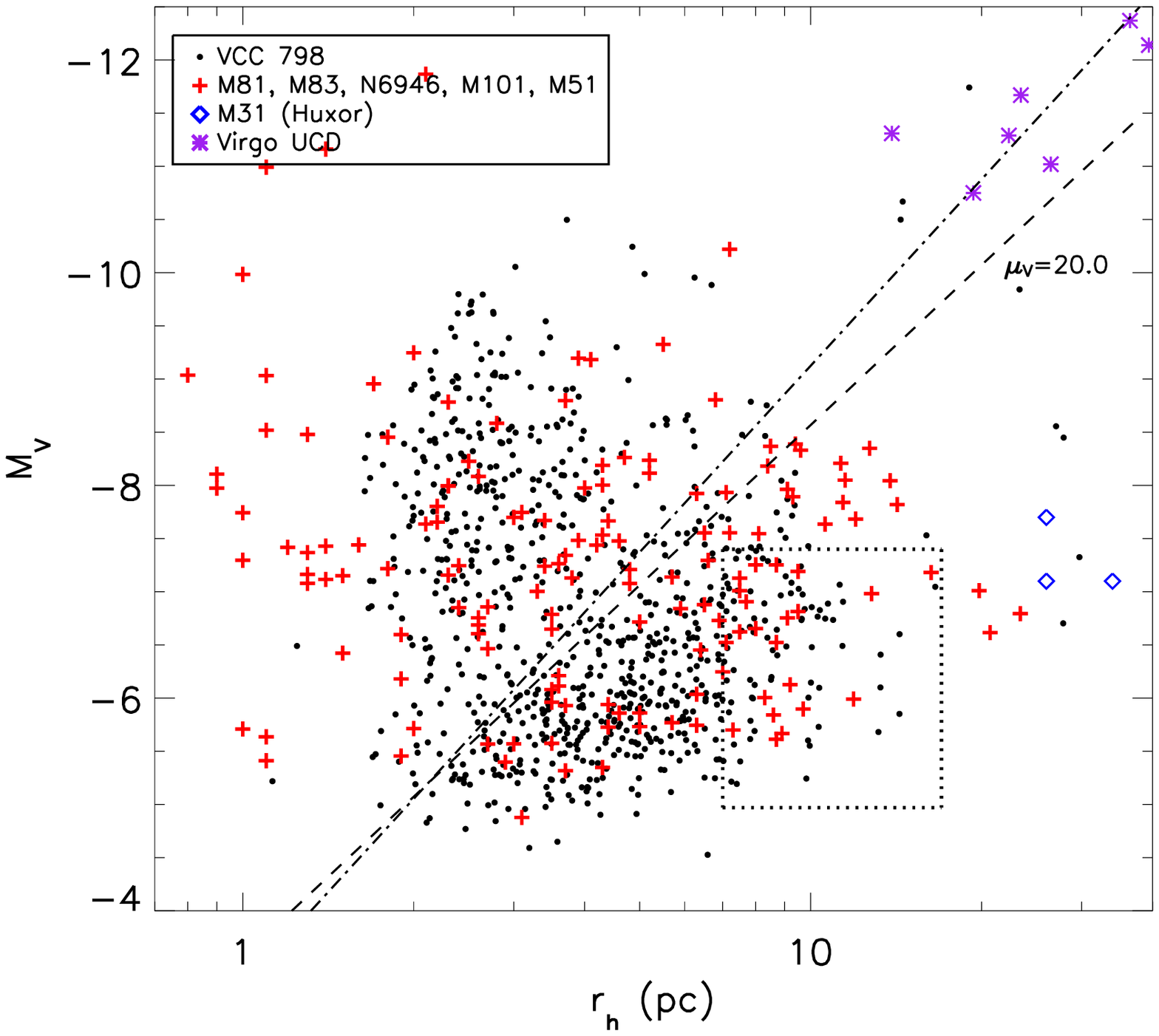}
\caption{Size-magnitude diagram for VCC~798 (dot) with old star
  cluster candidates (cross) in five spiral galaxies---M81, M83, NGC~6946,
  M101, M51 (Tables~3--7, Chandar \etal 2004).
  Photometry for VCC~798 and ``faint fuzzy'' selection box (dotted
  lines) is the
  same as for Figure~\ref{fig:mw_comp}. 
  We also show M31 extended star clusters from Huxor \etal (diamond)
  and UCDs from \hasegan \etal (asterisk).  The dashed line 
  represents a constant mean
  surface brightness, $\mu_V = 20\ {\rm mag\ arcsec}^{-2}$.  The dash-dotted
  line represents the mass-radius relation of $M\propto R^{2.33}$
  for star cluster complexes in M51 (Bastian \etal 2005).
  The spiral galaxies possess old
  star clusters that are similar in size and luminosity to
  the DSCs in our sample.
  \label{fig:diskgc_comp}}
\end{figure}

Because the DSCs appear to be associated with stellar {\it disks}, 
perhaps the most interesting comparison is with the old star cluster
populations of nearby spiral galaxies.  Chandar \etal (2004) conducted a
painstaking study of the old star cluster systems of five nearby
low-inclination spiral galaxies.  Using HST/WFPC2 multicolor imaging, they were
able to select a relatively clean sample of old star clusters in the
galaxies M81, M83, NGC~6946, M101, and M51, measuring magnitudes,
colors, and sizes for 145 GC candidates.  In
Figure~\ref{fig:diskgc_comp} we show the comparison between VCC~798
and the combined sample from these spiral galaxies in the
size-magnitude diagram (see Figure~8 in Chandar \etal (2004) for a
similar figure labeling star clusters from individual galaxies).  Most of
these spiral galaxies appear to possess star clusters that populate at
least a part of the DSC locus, with M81 and M51 having the largest
numbers.  It is possible that some of these objects are
background contaminants, but this is unlikely --- all of these spirals 
are at least 1.5~mag 
closer than the Virgo cluster, and the distribution of background
galaxies is constant in apparent magnitude.  Unlike the DSCs in our
sample, however, these star clusters are not predominantly red in
color but span the full range of star cluster colors.  This is perhaps
expected if the color of DSCs is correlated with the color of the
underlying stellar disk as these are spiral galaxies still in the act
of star formation.  Uncertainties in reddening are also likely to be
contributing to scatter in the colors.

The ``faint fuzzy'' star clusters discovered in NGC~1023 and 3384
are in many ways similar to the DSCs.  In 
Figures~\ref{fig:mw_comp}~and~\ref{fig:diskgc_comp}
we also show the rough selection region that Larsen \& Brodie (2000)
used to select ``faint fuzzy'' clusters.  In VCC~798, this selection box
contains a subset of the DSCs, selecting those that 
have larger half-light radii.  Given the red colors and old ages
of the ``faint fuzzies'' (Brodie \& Larsen 2002), it is likely that
these populations are the same phenomenon.  However, we point out that
not all of the
DSCs are faint (some have luminosities near the GCLF mean), and some
of them are as compact as GCs.  The defining characteristic of this
``excess'' star
cluster population is that they are at 
lower surface brightnesses than the bulk of
the globular clusters, hence we refer to this general population as
``diffuse'' to be as precise and inclusive as possible.

%%%%%%%%%%%%%%%%%%%%%%%%%%%%%%%%%%%%%%%%%%%%%%%%%%%%%%%%%%%%%%%%%%%%%%%%%%
%%%%%%%%%%%%%%%%%%%%%%%%%%%%%%%%%%%%%%%%%%%%%%%%%%%%%%%%%%%%%%%%%%%%%%%%%%

Recently, Huxor \etal (2005) discovered three stars clusters in M31 that
are both fairly luminous ($M_V\sim -7$) and extended ($r_h > 25$~pc).
When placed in the size-magnitude diagram (Figure~\ref{fig:mw_comp})
we find that while they would be some of the most extended star clusters
in VCC~798, they are not so different in size and magnitude from the
most extended and luminous DSCs.  However, the integrated colors and CMDs
of the M31 clusters show that they are blue and metal-poor, making them
distinctly different from most of the DSCs, 
and having perhaps more in common with
the Galactic halo GCs.  Their large galactocentric radii also suggest
that they belong to a halo population.

A recent investigation of the star cluster systems of nearby low
surface brightness dwarf galaxies by Sharina \etal (2005) also turned
up a number of low surface brightness star clusters that would meet
our criteria for DSCs.  In their investigation, a majority of the
globular cluster candidates in both dwarf spheroidal and dwarf
irregular (dI) galaxies can be considered DSCs.  The fraction appears
higher for the dI galaxies in their sample.  This is an interesting
complement to our findings because while we do not find many DSCs in the
ACSVCS dwarf elliptical galaxies (only VCC~9 has a significant
population) this could simply be due to low numbers in each galaxy and a
relatively high background.  The Sharina \etal sample is much closer
(2--6~Mpc) and so suffers much less from the problem of background galaxies.
The ACSVCS does
not contain dIs, but the presence of diffuse star clusters 
in dIs might be consistent with them residing in galactic disks.

\subsection{Diffuse Star Clusters in the Milky Way: Lost or Found?}

One of the notable claims made for diffuse star clusters is that there
are no known analogs in the Milky Way.  However, observations of other
nearby disk galaxies with HST as shown in the previous section 
indicate that such star clusters are
not uncommon.  This raises the question: if a population of old,
metal-rich, diffuse star clusters did exist in the disk of the Milky
Way, would we be able to detect them?  If so, how many might we expect
to find?  We can make a zero-order
estimate of the number of DSCs we might be expected to have found in the
Galaxy.  At the distance of the Virgo cluster, 
the ACS/WFC field of view covers $\sim261\ {\rm kpc}^2$ roughly
encompassing  a galactocentric radius of $\sim8$~kpc.  
If we view a galaxy nearly face on,
then we can uses this number to infer a DSC number surface density.
The closest analog to DSCs in
the Milky Way are the old open clusters in the disk of the Galaxy and we
can use their discovery efficiency to determine the likelihood of
finding DSCs.
Survey efficiency for old open clusters is difficult to quantify, 
but the observed spatial distribution 
roughly shows the area in the disk that has been most effectively
surveyed.  The distribution of old open clusters is most complete
between $l=160^\circ$ and $l=200^\circ$, 
and galactocentric radii of 7 to 13~kpc (see Friel 1995).
This gives an effective survey area in the disk of $42\ {\rm kpc}^2$.  The DSCs
are more luminous than the old open clusters but are of similar
surface brightness, so we expect that their detectability might be
roughly similar.  If we assume that the mean surface density of diffuse
star clusters in our images is an upper limit to that in an annulus
from 7--13~kpc,
then the number DSCs we might expect to see is a ratio of the areas
times the total number in the ACS/WFC field, or $0.16\times N_{DSC}$.  
This is the number of DSCs --- objects that are detected in our ACSVCS
images and classified as a diffuse star cluster --- that we would
expect to see in the Milky Way given the extent of old open
cluster surveys. 
If the Galaxy had a DSC population like that in VCC~798, then we might
expect to have detected $\lesssim25$~Milky Way DSCs in open cluster surveys.  
If the Galaxy has a smaller
number of diffuse clusters, like in VCC~1903, then we might only expect
to have found $\lesssim6$ star clusters.

We can also ask the question in reverse --- 
how many old open clusters are known in the disk of the Galaxy
that, if placed at the distance of the Virgo cluster and
observed in the ACSVCS, would have been classified as a diffuse star cluster?
Figure~\ref{fig:mw_comp} shows
that at least nine old open clusters in the Milky Way would have been
detected in the ACSVCS if they were in VCC~798.  While we realize that
old open clusters are not a perfect match for the DSCs and that we
have made many simplifying assumptions, 
these numbers raise two interesting possibilities: 1)
The fact that very few analogs to diffuse star clusters have been found in
the Galaxy does not necessarily mean that they do not exist.
Extrapolating from our survey, we expect that even if the Milky Way
had a moderate number of DSCs, only a few might be detectable in
surveys done to date, or 2)
The number of bright old open clusters may in fact be consistent with
the some of the less DSC-rich galaxies that we see.

Ongoing searches for new star
clusters with the Two Micron All Sky Survey are currently some of the most
promising avenues to discover highly obscured low surface brightness
star clusters (e.g.\ Bica, Dutra, \& Barbuy 2003), but these searches
are difficult due to variable foreground extinction and high stellar
densities in the Galactic disk.  A more complete census of disk star
clusters, especially for those with large radii, would be invaluable
for comparisons to diffuse cluster populations in other galaxies.

\subsection{Constraints on the Origin of the Diffuse Star Clusters}

Fellhauer \& Kroupa (2002) have proposed that diffuse star clusters
such as the ``faint fuzzies'' can be formed via the merging of star
cluster complexes, which are found in interacting galaxies.  Using
simulations, they show that star clusters that have the approximate
magnitudes and sizes of diffuse star clusters can be formed and survive
in a galactic disk for many Gyr.  The ultimate evolution of one of these merged
star cluster complexes ends up having a surface brightness and
luminosity similar to the diffuse clusters we see.  They also claim
that the same basic process can also form the ultracompact dwarf
galaxies (UCDs) that have been discovered in the Fornax and Virgo
clusters (Hilker \etal 1999; Drinkwater \etal 2003; \hasegan \etal 2005).  In
Figure~\ref{fig:diskgc_comp}, we plot the locations of the Virgo UCDs
and dwarf-globular transition objects from \hasegan \etal (2005, 2006).

Observationally, Bastian \etal (2005) observe that young star cluster
complexes in the disk of M51 are grouped hierarchically into complexes
which show evidence of merging.  Unlike star clusters, the complexes
show a mass-radius relation which goes as $M\propto R^{2.33\pm0.19}$,
which is close to a constant surface brightness (or mass density).
Although our data is not complete enough to determine a meaningful surface
brightness distribution of DSCs, the fact that the high surface
brightness envelope of diffuse star clusters appears to follow a line of nearly
constant surface brightness (or $M\propto R^2$) may be a clue to their
formation.  
Bastian \etal (2005) find that the mass-radius relation for giant
molecular clouds in M51 is also $M\propto R^2$.
In Figure~\ref{fig:diskgc_comp}, we show both a line of
constant surface brightness, $\mu_V=20\ {\rm mag\ arcsec}^{-2}$, and a
line following the mass-radius relation of Bastian \etal for star cluster
complexes which has been fixed to coincide with the Virgo UCDs.  While
it may be a coincidence, we note that this relation appears to
characterize both the UCDs and the upper envelope of the diffuse star
clusters.

Although our selection criteria for DSCs in the previous sections does
essentially involve a surface brightness cut, the constant surface
brightness upper envelope is not a product of our classification.  The
deficit of objects at this surface brightness shown for VCC~798 in
Figures~\ref{fig:sbhist} and \ref{fig:diskgc_comp} is independent our
our classification scheme.  For the fainter and more compact star
clusters, it is difficult to separate DSCs from GCs, but it is
especially for more extended and luminous star clusters that we see
the separation in surface brightness.  
In galaxies such as VCC~1535, the separation between the
DSCs and GCs is clearer.  We admit, however, that defining any sort of
mass-radius boundary with this data is likely to be uncertain.

We also note that there may more more tha one mechanism for the
production of diffuse star clusters.  VCC~1199 is a compact
elliptical galaxy that lies close in projection to M49, 
and whose cluster system is likely swamped by that of its giant
neighbor.  Using a simple model of the M49 GC system one expects that
100--200 GCs in the VCC~1199 field---$\sim50$--$100\%$ of total---are
associated with M49.  VCC~1192, another
satellite of M49, also has a high number of DSC candidates.  Although
M49 itself does not appear to have many DSCs, its outer halo appears to have
many more.  These clusters are not just the more extended GCs one
expects to find in the halo as they have the same properties as DSCs
in other galaxies.  Either they were accreted from other galaxies
(such as 1199 and 1192 or others), or perhaps the halo of M49 is more
hospitable to the formation and survival of diffuse star clusters.

\section{Conclusions}

The ability to obtain deep images of a large sample of nearby galaxies
has advanced the study of extragalactic star cluster systems to lower
surface brightness levels.  Some galaxies have a significant number of
diffuse star clusters which appear to be a population apart from
globular clusters.  Below, we summarize their properties and
some constraints on their origins.

\begin{enumerate}
\item {\it Structural Properties.}
  Diffuse star clusters are characterized by their low surface
  brightnesses, typically $\mu_g > 20$ and 
  $\mu_z > 19\ {\rm mag\ arcsec}^{-2}$.  Unlike GCs, whose sizes do not
  vary with their luminosity, the high surface brightness envelope of
  DSCs is roughly consistent with a constant surface brightness.  This
  hints that DSCs may be bounded by a mass-radius relationship of the
  type $M\propto R^\gamma$ where $\gamma\approx 2$.  
  The highest surface brightness DSCs are only slightly more diffuse
  than UCD-like objects.
  
\item {\it Diffuse star clusters are old or metal-rich, or both.}
  The diffuse star clusters tend to have red colors with
  $1.1 < g-z < 1.6$.  Their median colors are redder than the red
  globular cluster subpopulation, and they often match the color of the
  galaxy itself.  When compared to SSP models, these colors require that
  all of the DSC systems except those in VCC~9 have mean ages of older than
  5--13~Gyr or else have super-solar metallicities.  If we assume that
  they do not have super-solar abundances, then these diffuse clusters
  are long-lived, and any model for their survival must preserve them
  for as long as a Hubble time.
  
\item {\it Frequency of DSC systems}.
  We find twelve galaxies in our sample that contain a greater than
  $3\sigma$ excess of diffuse sources above what is expected from the
  background (Table~\ref{table:excesstable}).  If we restrict ourselves
  only to red diffuse sources with $(g$--$z) > 1.0$, then two more
  galaxies, VCC~1154 and VCC~2092 also pass this criterion.  All of
  these galaxies with the exceptions of VCC~1903, 9 and 1199 are
  morphologically classified as lenticulars.  The ACSVCS sample is
  magnitude limited for $M_B < -18.94$.  Nine of the 26 galaxies in this
  magnitude range (35\%) have a significant system of DSCs.  This is
  partially driven by the high fraction of S0s in this magnitude range.
  However, of the 38 ACSVCS galaxies with S0 classifications in the VCC,
  these nine are the only ones with significant DSC systems (24\% of S0s).  
  Given that of the 63 galaxies that were removed from the ACSVCS
  sample a large number were S0s, it is difficult to determine the true DSC
  frequency.  
  
\item {\it DSCs are spatially aligned with the galaxy light.}  
  In cases where a galaxy is highly inclined or the galaxy light is
  anisotropic within the ACS field of view, the DSCs visibly align
  themselves with the galaxy.  It is likely that the DSCs are a
  population of star clusters associated with the disks of galaxies,
  rather than with bulges or halos.  Although Larsen \& Brodie (2000)
  use deep WFPC2 data to claim that there is a deficit of ``faint
  fuzzies'' at the center of 
  NGC~1023, we are not able to make a similar statement about the DSC
  populations due to rising incompleteness in the high surface brightness
  central regions of some of our galaxies, and so we cannot comment on the
  frequency of ``ring''  structures such as those simulated by Burkert,
  Brodie, \& Larsen (2005). 

\item{\it The fraction of stellar luminosity contained in diffuse star 
  clusters for our DSC-excess sample is typically
  $1$--$7\times10^{-4}$.} VCC~9 and VCC~1199 are the exceptions.  It is
  possible that the lower luminosity galaxies in the sample also have
  DSCs, but if they form at similar efficiencies, then it would be
  difficult to notice them among the background contamination.  However,
  the central regions of the giant ellipticals are definitely deficient 
  in DSCs.

\item {\it Most DSC systems are associated with galactic disks, but many
  lenticulars do not host substantial DSC populations.}
  Both the spatial distributions of the diffuse clusters and the
  morphological classification of their host galaxies suggest that the
  formation and sustainability of DSCs is linked to the existence of a
  stellar disk.  Nine of the 12 galaxies with significant diffuse star cluster
  populations are morphologically classified as S0 (11 out of 14 if one
  counts VCC~1154 and 2092).  The existence of similar clusters in the
  spiral galaxies studied by Chandar \etal (2004) provides more evidence
  that disks are preferred environments for the formation and survival of
  diffuse star clusters.  However, not all disks harbor DSCs, and in fact
  the majority of the lenticulars in our sample do {\it not} have
  substantial diffuse star cluster systems.  
  One clue is that DSCs seem to exist preferentially in galaxies with
  visible dust---five of the nine S0s with substantial DSC populations
  also have visible dust at their centers.  Understanding the variations
  in DSC numbers between galaxies that are otherwise very similar will be
  important for determining the origin of diffuse star clusters.

\item{\it There may be more than one mode of diffuse cluster formation.}
  Two notable exceptions in our sample are not lenticular galaxies.  VCC~9
  is an isolated low surface brightness dwarf elliptical galaxy that has a
  central concentrated population of DSCs.  The existence of DSCs in the
  dwarf spheroidals of the Sharina \etal (2005) sample also suggest that
  DSCs can form outside of the disk environment (or that the dSphs previously
  had disks).  

\item {\it Environment does not appear to
  predict of the existence of a diffuse star cluster population.}
  We find that there is little trend between clustercentric radius or
  local galaxy density with the number of DSC candidates.  Although some
  of the galaxies with DSCs have nearby neighbors (VCC~798 and 1030),
  others are isolated.  The old ages of the DSCs also argue against an
  important role for very {\it recent} interactions.  Nevertheless, the fact
  that a VCC~798 and 1030 have close neighbors and that the existence
  of dust may be important are both signs that interactions may be one
  avenue for the formation of diffuse star clusters.
  
\item {\it The lack of many known DSC analogs in the Milky Way does not
  mean they do not exist.}  Given the difficulty of searching the
  Galactic plane for low surface brightness star clusters, we estimate
  that the Galaxy could have a DSC-like star cluster population and only
  a few would have been within the range of previous surveys.

\item {\it The old open clusters are the closest Galactic counterpart to
  the diffuse star clusters.}  These relatively metal-rich disk clusters
  are similar to the DSCs we observe, but are generally less luminous and
  probably younger.

\end{enumerate}

Future observations will further elucidate the origins of these diffuse
star cluster populations.  The ACS Fornax Survey will image 43
early-type galaxies to similar depths as the ACSVCS, but in a different
environment (\jordan \etal 2006).  
We are also obtaining deep optical and near-infrared
photometry of hundreds of star clusters in VCC~798 for the purpose of
constraining their ages, metallicities, and structural parameters.  

\acknowledgments

We thank Sidney van den Bergh for making his catalog of open clusters
available to us before publication.  We also thank Arunav Kundu for
sending us his photometry of M87 globular clusters.
Support for program GO-9401 was provided through a grant from the
Space Telescope Science Institute, which is operated by the
Association for Research in Astronomy, Inc., under NASA contract NAS5-26555.
Partial support for this work was provided by NASA LTSA grant
NAG5-11714 to PC.  M. J. W. acknowledges support
through NSF grant AST 02-05960.
This research has made use of the NASA/IPAC Extragalactic Database
(NED) which is operated by the Jet Propulsion Laboratory, California
Institute of Technology, under contract with the National Aeronautics
and Space Administration. 
This research has made use of the Vizier catalog service 
(Ochsenbein \etal 2000), which is hosted by the Centre de Donn{\'e}es
astronomiques de Strasbourg. 
 

%% copy editing.  Individual instruments can be provided in parentheses,
%% after the keyword, but they will not be verified.

Facilities: HST(ACS)

\end{document}